\documentclass[bibyear]{aa}
\usepackage{graphicx}

\def\HII{H{\sc ii} }

\def\UCHC{UC/HC~H{\sc ii}}

\def\kms{\mbox{km~s$^{-1}$}}
\def\cmc{cm$^{-3}$}

\def\mjy{~mJy~beam$^{-1}$}

\def\g35{G35.20N}

\begin{document}
\title{Binary system and jet precession and expansion in G35.20$-$0.74N} 
\author{M.\ T.\ Beltr\'an\inst{1}, R.\ Cesaroni\inst{1}, L.\ Moscadelli\inst{1}, \'A. S\'anchez-Monge\inst{2},
T.\ Hirota \inst{3, 4}, \and M.\ S.\ N.\ Kumar\inst{5, 6}
}
%
\institute{
INAF-Osservatorio Astrofisico di Arcetri, Largo E.\ Fermi 5,
I-50125 Firenze, Italy
\and
I.\ Physikalisches Institut, Universit\"at zu K\"oln, 
Z\"ulpicher Str.\ 77, D-50937 K\"oln, Germany
\and
National Astronomical Observatory of Japan, Mitaka, Tokyo 181-8588, JAPAN
\and
Department of Astronomical Sciences, SOKENDAI (The Graduate University for Advanced Studies ), Mitaka, Tokyo 181-8588,
JAPAN
\and
Centre for Astrophysics, University of Hertfordshire, College Lane, Hatfield, AL10 9AB, UK
\and
Instituto de Astrof\'{\i}sica e Ci\^encias do Espa\c{c}o, Universidade do Porto, CAUP, Rua das Estrelas, 4150-762, Porto, Portugal
}
\offprints{M.\ T.\ Beltr\'an, \email{mbeltran@arcetri.astro.it}}
\date{Received date; accepted date}

\titlerunning{Binary system and jet precession in G35.20$-$0.74N}
\authorrunning{Beltr\'an et al.}

\abstract
{Atacama Large Millimeter/submillimeter Array (ALMA) observations of the high-mass star-forming region G35.20$-$0.74N have
revealed the presence of a Keplerian disk in core B rotating about a massive
object of 18~$M_\odot$, as computed from the velocity field. The luminosity of such a massive star would be comparable to (or higher than) the
luminosity of the whole star-forming region. To solve this problem it has been
proposed that core B could harbor a binary system. This could also explain
the possible precession  of the radio jet associated with this core, which has
been suggested by its S-shaped morphology.}
{We establish the origin of the free-free emission from core B and investigate
the existence of a binary system at the center of this massive core and
the possible precession of the radio jet.}
{We carried out VLA continuum observations of G35.20$-$0.74N at 2~cm in the B configuration and at 1.3~cm
and 7~mm in the A and B configurations. The bandwidth at 7~mm covers the CH$_3$OH
maser line at 44.069~GHz. Continuum images at 6 and 3.6~cm in the A configuration
were obtained from the VLA archive. We also carried out VERA observations of the H$_2$O maser line at 22.235~GHz.}
{The observations have revealed the presence of a binary system of \UCHC\ regions
at the geometrical center of the radio jet in G35.20$-$0.74N. This binary system, which is 
associated with a Keplerian  rotating disk,  consists of two B-type stars of 11 and
6~$M_\odot$. The S-shaped morphology of the radio jet 
has been successfully explained as being due to
precession produced by the binary system. The analysis of
the precession of the radio jet has allowed us to better interpret the IR
emission in the region, which would be not tracing a wide-angle cavity open by a
single outflow with a position angle of $\sim$55$^\circ$, but two different
flows: a precessing one in the NE--SW direction associated with the radio jet, and a
second one in an almost E--W direction.    
Comparison of the radio jet images obtained at different epochs suggests that the
jet is expanding at a maximum speed on the plane of the sky of 300~\kms. The 
proper motions of the H$_2$O maser spots measured in the region also indicate
expansion in a direction similar to that of the radio jet. }
{We have revealed a binary system of high-mass young stellar objects embedded in the rotating disk in G35.20$-$0.74N. 
The presence of  a massive binary system  is in agreement with the theoretical predictions of high-mass star formation, 
 according to which the gravitational instabilities during the collapse
  would produce the fragmentation of the disk and the formation of such a system. For the first time, we have detected a high-mass young star associated with an
   \UCHC\
 region and at the same time powering a radio jet.}

\maketitle

\section{Introduction}

The discovery of circumstellar disks around early B-type stars (e.g., Beltr\'an
\& de Wit~\cite{beltran15} and references therein) in recent years has led to a
big leap in our understanding of high-mass star formation. It shows that B-stars
form via disk-mediated accretion confirming the predictions of theoretical
models (Bonnell \& Bate~\cite{bonnell06}; Keto~\cite{keto07}; Krumholz et
al.~\cite{krumholz09}). But theoretical predictions go further and suggest that 
gravitational instabilities during the collapse would cause the disk to
fragment, resulting in massive binary, triple or multiple systems, depending on
the initial mass of the cloud (Krumholz et al.~\cite{krumholz09}). Young
high-mass binary or multiple systems have been very elusive till now with only a
few convincing candidates (e.g., Shepherd et
al.~\cite{shepherd01}).  Next, although molecular outflows are ubiquitously
found in both low- and high-mass stars,  thermal ionized jets ---
common in low-mass protostars --- have so far been found in association
with only a handful of outflows driven by high-mass young stellar objects
(YSOs)  (e.g., Guzm\'an et al.~\cite{guzman12}). What is more, no thermal radio
jet known to be driven by a high-mass YSO
associated with an \HII region has ever been found.

\begin{table*}
\caption[] {Parameters of the maps$^{\rm a}$}
\label{table_parobs}
\begin{tabular}{ccccccc}
\hline
&&&\multicolumn{2}{c}{Synthesized beam}  
\\
\cline{4-5}
&\multicolumn{1}{c}{Wavelength} &
&\multicolumn{1}{c}{$HPBW$} &
\multicolumn{1}{c}{P.A.} &
\multicolumn{1}{c}{rms} 
\\
\multicolumn{1}{c}{Configuration} &
\multicolumn{1}{c}{(cm)} &
\multicolumn{1}{c}{Epoch} &
\multicolumn{1}{c}{(arcsec)} &
\multicolumn{1}{c}{(deg)} &  
\multicolumn{1}{c}{(\mjy)} 
\\
\hline
B &2.0    &2013 Oct. 18  &$0.44\times0.41$  &40     &0.033 \\
  &1.3     &2013 Dec. 6  &$0.25\times0.24$  &$-$35  &0.033 \\
  &0.7     &2013 Dec. 6  &$0.14\times0.12$  &$-$45  &0.040 \\
A &6.0    &1999  Sep. 15 &$0.54\times0.40$  &52     &0.033 \\
  &3.6     &1999 Sep. 15  &$0.24\times0.24$  &59     &0.027 \\
  &1.3     &2014 Jan. 24  &$0.08\times0.08$  &$-$46  &0.023 \\
  &0.7     &2014 Jan. 24  &$0.05\times0.04$  &$-$52  &0.033 \\
\hline
\end{tabular}

$^a$ Maps created with the robust parameter of Briggs (1995) set equal to 0. 
\end{table*}

In a progressive effort to understand massive star formation and verify the
predictions of theoretical models, we performed Atacama Large Millimeter/submillimeter Array (ALMA) observations of star-forming regions hosting B-type stars. One of our targets
was the well-known high-mass YSO G35.20$-$0.74N (hereafter \g35), which is
located at 2.2~kpc (Zhang et al.~\cite{zhang09}) and has a luminosity of
$\sim3\times10^4\,L_\odot$. This star-forming region is characterized by the
presence of a butterfly-shaped nebula, imaged at IR wavelengths, roughly
coinciding with a poorly collimated wide-angle bipolar outflow oriented NE--SW
(Gibb et al.~\cite{gibb03}; Qiu et al.~\cite{qiu13}).  Radio and IR images
reveal the presence of a collimated N--S jet (S\'anchez-Monge et
al.~\cite{sanchez-monge13}, \cite{sanchez-monge14}) originating from the same
center as the bipolar nebula and lying along the western border of the
wide-angle bipolar nebula/outflow. It has been suggested that the jet could be
precessing around the NE--SW direction, and be responsible for the
bipolar structure.

The ALMA observations have uncovered at least five cores across the waist of this
bipolar nebula (S\'anchez-Monge et al.~\cite{sanchez-monge14}), one of which
(core~B) appears to drive the N--S jet (S\'anchez-Monge et
al.~\cite{sanchez-monge13}). This core is spatially elongated and shows a velocity
gradient perpendicular to the axis of the bipolar nebula. S\'anchez-Monge et
al.~(\cite{sanchez-monge13}) demonstrate that the velocity field of the core can be
fitted with a Keplerian disk rotating about a massive object of 18~$M_\odot$. Such
a mass is unlikely to be concentrated in a single star because the luminosity of
an 18~$M_\odot$ star is comparable to the bolometric luminosity of the entire
\g35 star-forming region. To solve this luminosity problem, it has been argued that the luminosity of the region could have been 
underestimated if part of the stellar radiation had escaped through the outflow cavities (Zhang et al.~\cite{zhang13}).
Alternatively, the measured luminosity ($3\times10^4~L_\odot$) could arise from more than one
lower luminosity/mass star. This solution led S\'anchez-Monge et
al.~(\cite{sanchez-monge13}) to argue that the 18~$M_\odot$ unresolved object
consists of a binary system, with individual luminosities that can be as low as
$\sim7000~L_\odot$. The existence of a binary could cause precession of the
jet about the NE--SW axis of the disk, consistent with the scenario depicted above
for the origin of the bipolar nebula.

The next intriguing feature of core~B is the presence of an unresolved free-free
emission source inside this core. This source seems to be part of the N--S radio
jet, but has a spectral index $\alpha$ $>1.3$ ($S_\nu\propto\nu^\alpha$) between
6 and 3.6~cm (Gibb et al.~\cite{gibb03}), and $\sim$1.8 between 3.6 and 1.3~cm
(Codella et al.~\cite{codella10}), compatible with optically thick free-free
emission from an \HII region ($\alpha\sim2$). In contrast, thermal radio jets
have typically weak continuum fluxes of a few mJy and $\alpha=1.3-0.7/\epsilon
\lesssim 1.3$, where $\epsilon$ is related to the collimation of the jet (for a
prototypical biconical  jet $\epsilon=1$ and $\alpha$= 0.6:
Reynolds~\cite{reynolds86}; Anglada~\cite{anglada96}). While the \HII region
scenario appears to be compatible with the measured spectral indices, it cannot
rule out the possibility that the free-free emission from core B originates in
the N--S thermal jet because the observations at different wavelengths
have been performed with different angular resolutions. This makes it difficult to reliably
estimate the spectral index (3.6~cm and 6~cm: $\sim 0\farcs2$--$0\farcs5$;
1.3~cm: $\sim 1\farcs2$). 


To establish the origin of the free-free emission from core B and test the
hypothesis of a binary system in this core, we performed deep 2~cm, 1.3~cm, and 7~mm
continuum observations of \g35 with the Karl G.\ Jansky Very Large Array 
(VLA).

\begin{figure*}
\vspace{-6.5cm}
\centerline{
\hspace{5cm}
\includegraphics[angle=0,width=14cm]{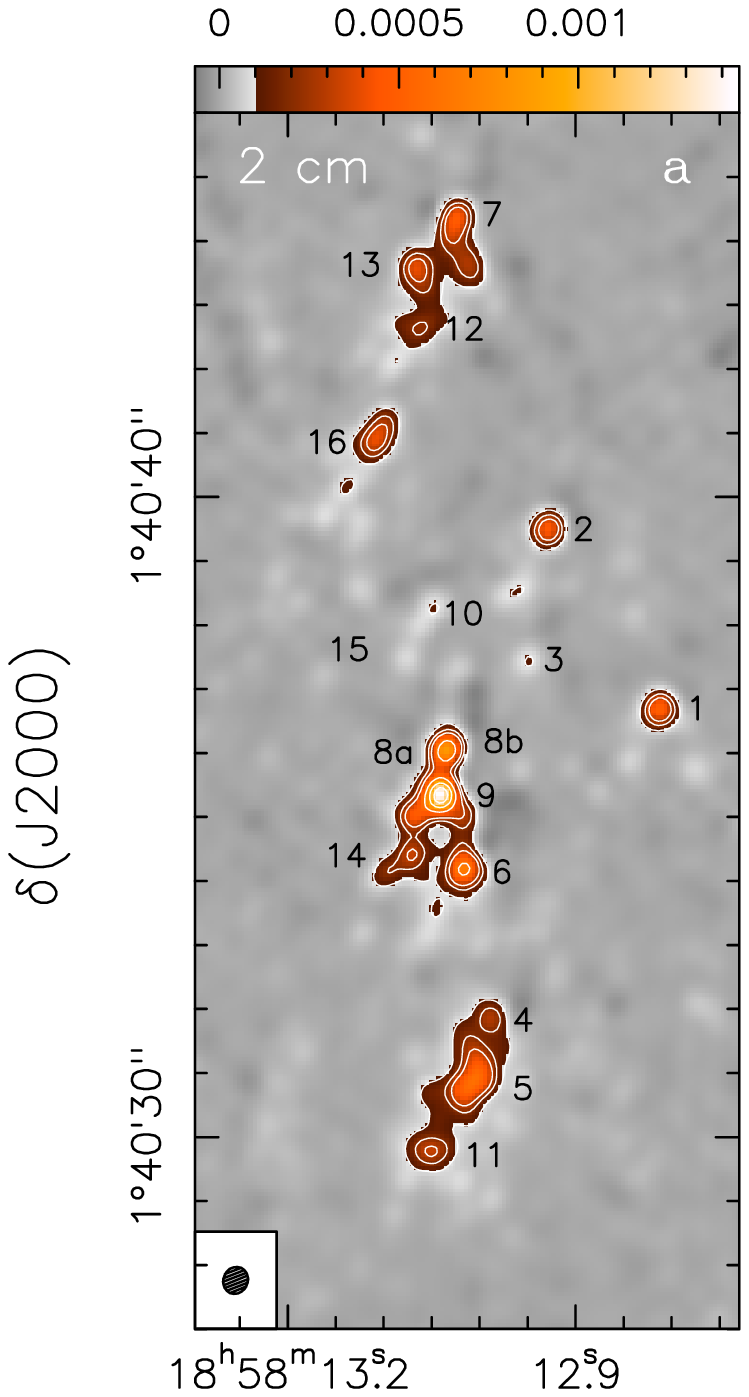}
\hspace{-9.575cm}
\includegraphics[angle=0,width=14cm]{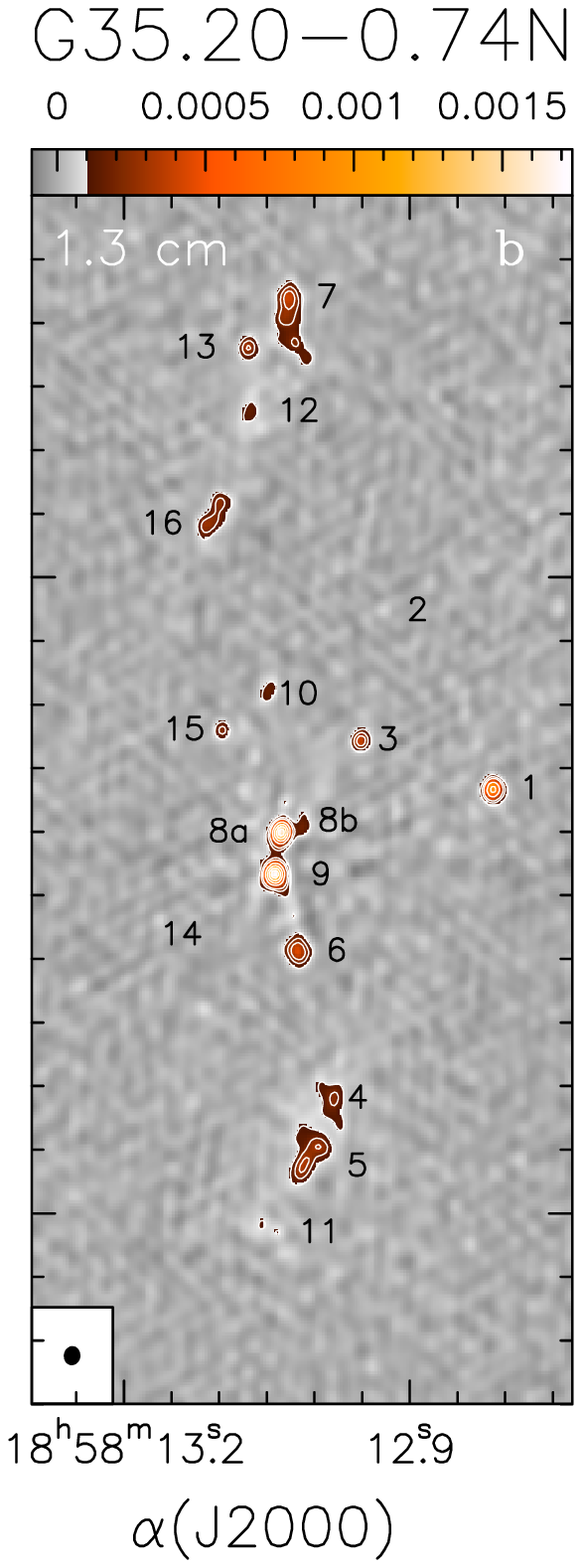}
\hspace{-9.575cm}
\includegraphics[angle=0,width=14cm]{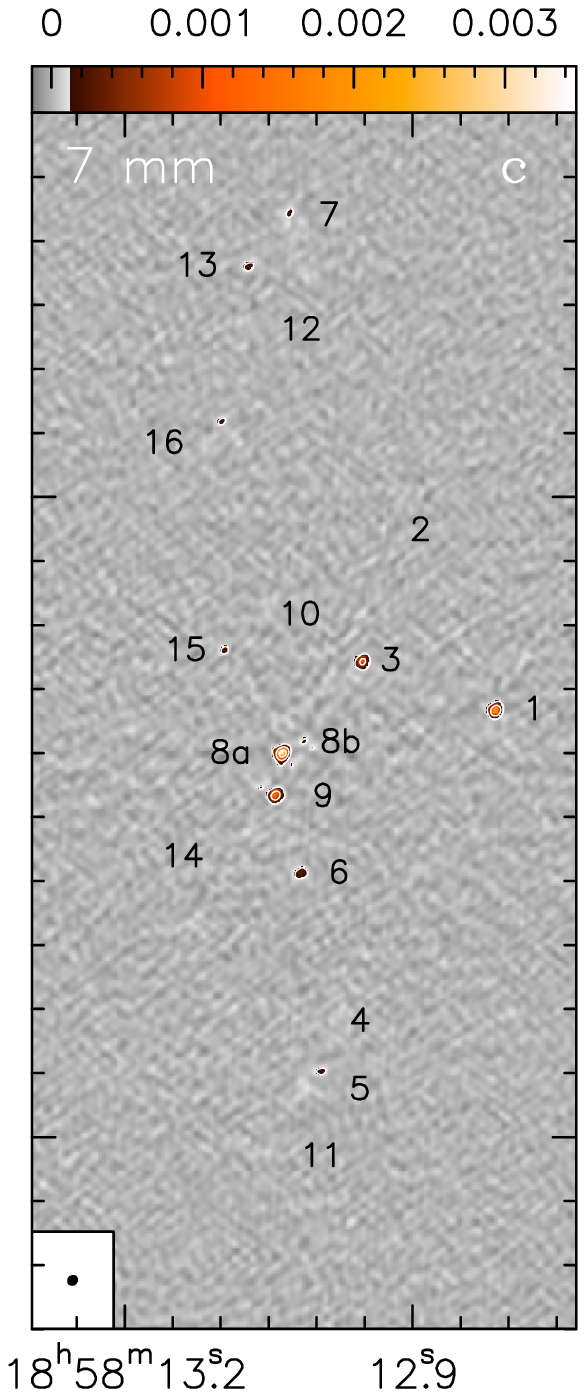}}
\caption{VLA map of the 2\,cm ({\it a}), 1.3\,cm ({\it b}), and 7\,mm ({\it c})
continuum emission of \g35 observed in B configuration. The VLA synthesized beam is shown in the lower left
corner of each panel. The contours are 3, 6, 9, 18, 27, 36, and 45 times 1\,$\sigma$,
which is 0.033\mjy, for ({\it a}) and ({\it b}), and 3, 18, and  54 times 1\,$\sigma$,
which is 0.04\mjy, for ({\it c}). The numbers correspond to the centimeter continuum
sources 
detected (see Table~\ref{table_cont1}).}
\label{fig-cm}
\end{figure*}

\section{Observations}

\subsection{VLA observations}

We carried out interferometric observations with the VLA of the
NRAO\footnote{The National Radio Astronomy Observatory is a facility of the
National Science Foundation operated under cooperative agreement by Associated
Universities, Inc.} in the B configuration at 2~cm on October 18, 2013, and at
1.3~cm and   7~mm on December 6, 2013, and in the A configuration at  1.3~cm
and   7~mm on February 24, 2014 (project 13B-033). The field phase center was 
$\alpha$(J2000)=18$^{\rm h}$\,58$^{\rm m}$\,13$\fs$030, 
$\delta$(J2000)=$+01\degr$\,40$'$\,36$\farcs$00. 
The quasar J1331+305 (3C286) was used as absolute
amplitude calibrator and J1851+0035 as
phase calibrator. The observations at 2\,cm were
performed with the 8-bit samplers while those at 1.3\,cm and 7\,mm with the
3-bit samplers.  The total 2\,GHz bandwidth at 2\,cm was covered with
18 spectral windows of 128 1\,MHz channels, while the 8\,GHz bandwidth at
1.3\,cm and 7\,mm was covered with 64 spectral windows of 128 1\,MHz 
channels. To construct the continuum maps, we averaged all the channels and the
integration time with an averaging interval of 10 seconds for the B
configuration and 5 seconds for the A configuration. 

The data were calibrated and imaged using the CASA\footnote{The CASA package is available at
http://casa.nrao.edu/}
software package. Further imaging and analysis was done with the GILDAS\footnote{The GILDAS package is available at
http://www.iram.fr/
IRAMFR/GILDAS}
software package. The final maps were created with the Briggs robustness parameter (Briggs~\cite{briggs95})
set equal to 0. Resolution and sensitivity of the observations at the different
wavelengths and configurations are given in Table~\ref{table_parobs}. 

The bandwidth at 1.3\,cm covers the H$_2$O maser line at
22.235\,GHz, while that at 7\,mm covers the CH$_3$OH maser
line at 44.069\,GHz. The spectral resolution of the 1.3\,cm
observations is $\sim$13.4\,\kms, insufficient to properly
study the water maser emission, which will not be discussed further in this
work. The spectral resolution of the 7\,mm observations is 7\,\kms. 

Continuum images at 6 and 3.6 cm in the A configuration were obtained from the
VLA archive (project AG0576). The observations were carried out on September 15,
1999, with phase center $\alpha$(J2000)=18$^{\rm h}$\,58$^{\rm m}$\,12$\fs$93,
$\delta$(J2000)=$+01\degr$\,40$'$\,36$\farcs$5. The final maps have been
reprojected to the phase center of our VLA observations. The data were
re-calibrated and imaged with CASA to create maps with the same Briggs
robustness parameter as the 2 and 1.3~cm and 7~mm maps. The maps and the fluxes
are very similar to those obtained by Gibb et al.~(\cite{gibb03}) with uniform
weighting.

\begin{sidewaystable*}
\caption[] {Position, peak intensity, and integrated flux density of the cores
estimated from the B-array images.}
\label{table_cont1}
\begin{tabular}{cccccccccccc}
\hline
&&\multicolumn{2}{c}{Peak Position$^a$}
\\
 \cline{3-4} 
\multicolumn{1}{c}{Our}
&\multicolumn{1}{c}{Gibb's}
&\multicolumn{1}{c}{$\alpha({\rm J2000})$} &
\multicolumn{1}{c}{$\delta({\rm J2000})$} &
&\multicolumn{1}{c}{$I^{{\rm peak}, b}_{\rm 2\,cm}$} &
\multicolumn{1}{c}{$S_{\rm 2\,cm}^b$} &
\multicolumn{1}{c}{$I^{{\rm peak}, b}_{\rm 1.3\,cm}$} &
\multicolumn{1}{c}{$S_{\rm 1.3\,cm}^b$} &
\multicolumn{1}{c}{$I^{{\rm peak}, b}_{\rm 7\,mm}$} &
\multicolumn{1}{c}{$S_{\rm 7\,mm}^b$} &
\\
\multicolumn{1}{c}{numbering} &
\multicolumn{1}{c}{numbering} &
\multicolumn{1}{c}{h m s}&
\multicolumn{1}{c}{$\degr$ $\arcmin$ $\arcsec$} &
&\multicolumn{1}{c}{(mJy/beam)} & 
\multicolumn{1}{c}{(mJy)} &
\multicolumn{1}{c}{(mJy/beam)} & 
\multicolumn{1}{c}{(mJy)} &
\multicolumn{1}{c}{(mJy/beam)} & 
\multicolumn{1}{c}{(mJy)} &
\\
\hline
1      &6  &18 58 12.814 &01 40 36.68    &&0.55$\pm$0.03  &0.55$\pm$0.13    &0.97$\pm$0.03   &1.00$\pm$0.25   &1.97$\pm$0.04  &1.97$\pm$0.51   \\
2$^c$  &5  &18 58 12.927 &01 40 39.48    &&0.49$\pm$0.03  &0.49$\pm$0.11    &$<$0.10	      &$<$0.10	     &$<$0.12	      &$<$0.12         \\
3      &-  &18 58 12.953 &01 40 37.44    &&0.18$\pm$0.03  &0.18$\pm$0.03    &0.39$\pm$0.03   &0.39$\pm$0.09   &1.00$\pm$0.04  &1.00$\pm$0.25   \\ 
4      &9  &18 58 12.980 &01 40 31.81    &&0.32$\pm$0.03  &0.40$\pm$0.06    &0.23$\pm$0.03   &0.40$\pm$0.04   &$<$0.12        &$<$0.12         \\
5      &10 &18 58 12.995 &01 40 31.04    &&0.60$\pm$0.03  &1.60 $\pm$0.13   &0.32$\pm$0.03   &0.90$\pm$0.07   &0.17$\pm$0.04  &0.17$\pm$0.16  \\
6      &-  &18 58 13.017 &01 40 34.12    &&0.74$\pm$0.03  &0.90$\pm$0.18    &0.47$\pm$0.03   &0.60$\pm$0.11   &0.25$\pm$0.04 &0.30$\pm$0.04  \\
7      &1  &18 58 13.027 &01 40 44.40    &&0.52$\pm$0.03  &1.10 $\pm$0.11   &0.40$\pm$0.03   &1.10$\pm$0.07   &0.19$\pm$0.04 &0.19$\pm$0.04  \\
8      &7  &18 58 13.036 &01 40 36.00    &&0.85$\pm$0.03  &0.85$\pm$0.22       \\      
8a     &   &18 58 13.036 &01 40 36.00		                            &&&&1.69$\pm$0.03  &1.80$\pm$0.41 &3.47$\pm$0.04   &3.50$\pm$0.90   \\
8b$^d$ &   &18 58 13.014 &01 40 36.18					    &&&&0.13$\pm$0.03  &0.13$\pm$0.03 &0.15$\pm$0.04   &0.15$\pm$0.04 \\
9      &8  &18 58 13.041 &01 40 35.32    &&1.48$\pm$0.03  &2.30 $\pm$0.33   &1.75$\pm$0.03   &2.00$\pm$0.43   &1.50$\pm$0.04 &1.50$\pm$0.36    \\ 
10     &-  &18 58 13.049 &01 40 38.16    &&0.17$\pm$0.03  &0.17$\pm$0.03    &0.15$\pm$0.03   &0.15$\pm$0.03    &$<$0.12 	  &$<$0.12	   \\
11     &11 &18 58 13.054 &01 40 29.80    &&0.34$\pm$0.03  &0.40$\pm$0.06    &0.13$\pm$0.03   &0.13$\pm$0.03   &$<$0.12  	  &$<$0.12	  \\
12     &2+3  &18 58 13.067 &01 40 42.61  &&0.25$\pm$0.03  &0.30$\pm$0.03    &0.14$\pm$0.03   &0.14$\pm$0.03    &$<$0.12     &$<$0.12	  \\
13     &-  &18 58 13.070 &01 40 43.61    &&0.45$\pm$0.03  &0.70$\pm$0.09    &0.32$\pm$0.03   &0.32$\pm$0.07   &0.22$\pm$0.04 &0.22$\pm$0.04  \\
14$^c$ &-  &18 58 13.070 &01 40 34.38    &&0.28$\pm$0.03  &0.40$\pm$0.05    &$<$0.10	       &$<$0.10 	  &$<$0.12	   &$<$0.12	   \\
15     &-  &18 58 13.097 &01 40 37.60    &&$<$0.10        &$<$0.10	    &0.24$\pm$0.03   &0.24$\pm$0.04   &0.23$\pm$0.04   &0.23$\pm$0.04  \\ 
16     &-  &18 58 13.110 &01 40 40.84    &&0.42$\pm$0.03  &0.60$\pm$0.09    &0.28$\pm$0.03   &0.60$\pm$0.05   &0.18$\pm$0.04  &0.18$\pm$0.04  \\
\hline
    
\end{tabular}

  $^a$ Position estimated from the 1.3\,cm map. \\
  $^b$ Peak intensity and integrated flux density corrected for primary beam
  response.  \\
  $^c$ Position estimated from the 2\,cm map.\\
  $^d$ Position estimated from the 7\,mm map.\\
  \\
\end{sidewaystable*}

\begin{sidewaystable*}
\caption[]{Peak intensity and flux density of the cores estimated from the
A-array images.}
\label{table_cont2}
\begin{tabular}{ccccccccccc}
\hline
\multicolumn{1}{c}{Our}&
\multicolumn{1}{c}{Gibb's}&
\multicolumn{1}{c}{$I^{{\rm peak}, a}_{\rm 6\,cm}$} &
\multicolumn{1}{c}{$S_{\rm 6\,cm}^a$} &
\multicolumn{1}{c}{$I^{{\rm peak}, a}_{\rm 3.6\,cm}$} &
\multicolumn{1}{c}{$S_{\rm 3.6\,cm}^a$} &
\multicolumn{1}{c}{$I^{{\rm peak}, a}_{\rm 1.3\,cm}$} &
\multicolumn{1}{c}{$S_{\rm 1.3\,cm}^a$} &
\multicolumn{1}{c}{$I^{{\rm peak}, a}_{\rm 7\,mm}$} &
\multicolumn{1}{c}{$S_{\rm 7\,mm}^a$} &
\\
\multicolumn{1}{c}{numbering} &
\multicolumn{1}{c}{numbering} &
\multicolumn{1}{c}{(mJy/beam)} & 
\multicolumn{1}{c}{(mJy)} &
\multicolumn{1}{c}{(mJy/beam)} & 
\multicolumn{1}{c}{(mJy)} &
\multicolumn{1}{c}{(mJy/beam)} & 
\multicolumn{1}{c}{(mJy)} &
\multicolumn{1}{c}{(mJy/beam)} & 
\multicolumn{1}{c}{(mJy)} &
\\
\hline
1   &6    &0.16$\pm$0.03 &0.16$\pm$0.05 &0.27$\pm$0.03 &0.30$\pm$0.07 &0.95$\pm$0.02 &0.95$\pm$0.24	&1.79$\pm$0.03      &1.90$\pm$0.46   \\
2   &5    &0.13$\pm$0.03 &0.13$\pm$0.04 &0.20$\pm$0.03 &0.20$\pm$0.05 &0.22$\pm$0.02 &0.22$\pm$0.04	&0.12$\pm$0.03      &0.12$\pm$0.03  \\
3   &-    &$<$0.10	 &$<$0.10	&0.12$\pm$0.03 &0.12$\pm$0.03 &0.35$\pm$0.02 &0.35$\pm$0.08	&0.94$\pm$0.03      &1.00$\pm$0.23  \\ 
4   &9    &0.75$\pm$0.03 &1.60$\pm$0.19 &0.35$\pm$0.03 &1.00$\pm$0.08 &$<$0.07  	      &$<$0.07  	&$<$0.10	    &$<$0.10	 \\
5   &10   &0.43$\pm$0.03 &0.90$\pm$0.12 &0.26$\pm$0.03 &0.70$\pm$0.05 &0.14$\pm$0.02 &0.20$\pm$0.02	&$<$0.10	    &$<$0.10	 \\
6   &-    &$<$0.10	 &$<$0.10	&$<$0.08       &$<$0.08       &0.17$\pm$0.02 &0.20$\pm$0.03	&$<$0.10	    &$<$0.10	 \\
7   &1    &0.70$\pm$0.03 &1.20$\pm$0.17 &0.43$\pm$0.03 &1.10$\pm$0.10 &0.14$\pm$0.02 &0.14$\pm$0.02	&$<$0.10	    &$<$0.10	 \\
8a  &7	  &0.24$\pm$0.03 &0.30$\pm$0.07 &0.43$\pm$0.03 &0.60$\pm$0.12 &1.33 $\pm$0.02	      &1.50 $\pm$0.34	&2.28$\pm$0.03      &3.10$\pm$0.60 \\
8b  &	  &0.27$\pm$0.03 &0.27$\pm$0.07 &0.10$\pm$0.03 &0.10$\pm$0.03 &0.12$\pm$0.02 &0.12$\pm$0.02	&0.16$\pm$0.03      &0.16$\pm$0.03   \\
9   &8    &0.54$\pm$0.03 &0.70$\pm$0.14 &0.30$\pm$0.03 &0.60$\pm$0.08 &1.22 $\pm$0.02	      &1.70 $\pm$0.30	&0.82$\pm$0.03      &1.10$\pm$0.20     \\ 
10  &-    &$<$0.10	 &$<$0.10	&$<$0.08       &$<$0.08       &$<$0.07  	     &$<$0.07	       &$<$0.10 	    &$<$0.10	 \\
11  &11   &0.90$\pm$0.03 &0.90$\pm$0.11 &0.21$\pm$0.03 &0.80$\pm$0.04 &$<$0.07  	     &$<$0.07	       &$<$0.10 	    &$<$0.10	 \\
12  &2+3  &0.90$\pm$0.03 &0.90$\pm$0.16 &0.38$\pm$0.03 &1.40$\pm$0.09 &$<$0.07  	     &$<$0.07	       &$<$0.10 	    &$<$0.10	 \\
13  &-    &$<$0.10	 &$<$0.10	&$<$0.08       &$<$0.08	    &0.19$\pm$0.02 &0.19$\pm$0.04     &0.11$\pm$0.03	  &0.11$\pm$0.03  \\
14  &-    &$<$0.10	 &$<$0.10	&$<$0.08       &$<$0.08	    &$<$0.07		    &$<$0.07	      &$<$0.10  	  &$<$0.10     \\
15  &-    &$<$0.10	 &$<$0.10	&$<$0.08       &$<$0.08	    &$<$0.07		    &$<$0.07	      &$<$0.10  	  &$<$0.10     \\ 
16  &-    &$<$0.10	 &$<$0.10	&$<$0.08       &$<$0.08	    &0.11$\pm$0.02 &0.20$\pm$0.04     &$<$0.10  	  &$<$0.10     \\
\hline
    
\end{tabular}

  $^a$ Peak intensity and integrated flux density corrected for primary beam
  response.  \\
\\
\end{sidewaystable*}

\subsection{VERA observations}

VLBI Exploration of Radio Astrometry (VERA) observations of the H$_{2}$O
maser line at 22.235~GHz were carried out on September 06, October 10,
November 05, December 03, 2013, and January 10, 2014. All four stations of
VERA were used in all sessions, providing a maximum baseline length of
2270~km. Observations were made in the dual beam phase-referencing mode;
G35.20N and a reference source J1858+0313
($\alpha$(J2000)=18$^{\rm h}$ 58$^{\rm m}$ 02$\fs$352576,
$\delta$(J2000)=+03$^{\circ}$ 13$'$ 16$\farcs$30172) were observed
simultaneously with the separation angle of 1.55$^\circ$. The
instrumental phase difference between the two beams was measured
continuously during the observations by injecting artificial noise
sources into both beams at each station (Honma et al.~\cite{honma2008a}).

The data were recorded onto magnetic tapes sampled with 2-bit
quantization at a rate of 1024~Mbps. Among a total bandwidth of 256~MHz,
one IF channel and the rest of 15 IF channels with a 16~MHz bandwidth
each were assigned to G35.20N and J1858+0313, respectively. A bright
extragalactic radio source, J1743-0350, was observed every 80 minutes as
a delay and bandpass calibrator.

Correlation processing was carried out on the Mitaka FX correlator
located at the NAOJ Mitaka campus. For the H$_{2}$O maser line, the
spectral resolution was set to be 15.625~kHz, corresponding to the
velocity resolution of 0.21~km~s$^{-1}$.

Data reduction was performed using the NRAO AIPS. First, results of the
dual-beam phase calibration and the correction for the approximate delay
model adopted in the correlation processing were applied to the
visibility data (Honma et al.~\cite{honma2008b}). Next, instrumental delays and phase
offsets among all of the IF channels were removed by the AIPS task FRING
on J1743-0350. Finally, residual phases were calibrated by the AIPS task
FRING on J1858+0313. These FRING solutions were applied to the target
source G35.20N. Synthesis imaging and deconvolution (CLEAN) were
performed using the AIPS task IMAGR. The uniform weighted synthesized
beam size (FWMH) was typically 1.5~mas$\times$0.8~mas with a position
angle of $-$40$^\circ$. The peak positions and flux densities of maser
features were derived by fitting elliptical Gaussian brightness
distributions to each spectral channel map using the AIPS task SAD.

\begin{figure*}
\centerline{\includegraphics[angle=-90,width=18cm]{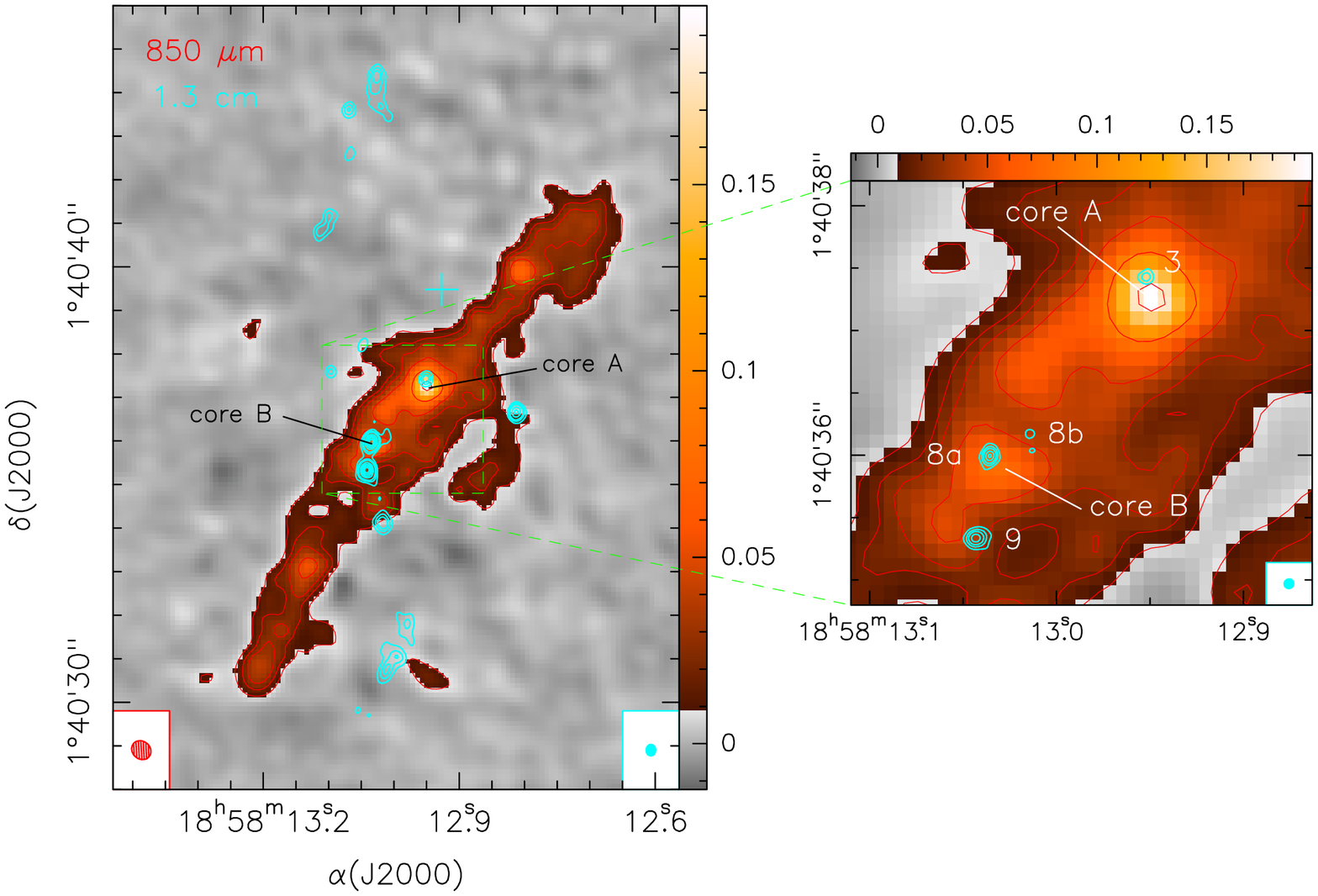}}
\caption{({\it Left panel}) Overlay of the ALMA 870\,$\mu$m continuum emission ({\it
colors and red contours}) from S\'anchez-Monge et al.~(\cite{sanchez-monge13}) on the 
VLA 1.3\,cm continuum emission ({\it cyan contours}) observed in B configuration. The cyan
cross indicates the position of source 2, not detected at 1.3\,cm, as
estimated from the
2\,cm map. The
ALMA and VLA synthesized beam are shown in the lower left and lower right 
corner, respectively. Red contours are 5, 10, 15, 25, 50, and 100 times 1\,$\sigma$, which
is 1.8\mjy. Cyan contours are the same as in Fig.~\ref{fig-cm}. ({\it Right panel})
Close-up of the central region around cores A and B that shows the ALMA 870\,$\mu$m continuum emission 
image overlaid with the VLA 1.3\,cm continuum emission ({\it cyan
contours}) observed in A configuration. Cyan contours are 3, 6, 18, 36, and 54 times 
1\,$\sigma$, which is 0.023\mjy. Red contours are the same as in the {\it left panel}.  
The VLA synthesized beam is shown in the lower right 
corner. The numbers correspond to the centimeter continuum sources in 
Table~\ref{table_cont2}.}
\label{fig-alma}
\end{figure*}

\section{Results}

\subsection{Continuum emission}
\label{cont-em}

Figure~\ref{fig-cm} shows the maps of the continuum emission at 2 and 1.3\,cm,
and 7\,mm towards \g35 observed in the B configuration with the VLA. A total of
16 sources have been detected in at least one wavelength. The positions
of the sources and their peak intensities and integrated flux densities estimated
in the B and A configuration are given in Tables~\ref{table_cont1} and
\ref{table_cont2}, respectively. For comparison, the tables also indicate the
number assigned by Gibb et al.~(\cite{gibb03}) to the radio sources. Because there are
far fewer sources detected at 6 and 3.6\,cm than we detected at  2\,cm, 1.3\,cm, and 7\,mm, we preferred to 
re-number the radio sources ordering them by increasing right ascension.
Table~\ref{table_cont2} gives also the peak intensity and integrated flux
density at 6 and 3.6\,cm.   

Source 2 shows a peculiar behavior at 1.3\,cm because it is clearly detected in the A configuration (see Table~\ref{table_cont2}) but not in 
the B configuration (its intensity is $<$0.1\,\mjy, a factor $\sim$2.2 lower than in the A configuration). A possible explanation 
could be source variability. In this case the timescales for this variability should be very short (1.5\,months), but Hofner et al.~(\cite{hofner07}) have 
reported strong variability on timescales of only one day for the source I20var in IRAS 20126+4104.    

As seen in Tables~\ref{table_cont1} and \ref{table_cont2}, sources 10, 11, 12, and 15 are unresolved at 1.3\,cm and 7\,mm
in the B configuration, as indicated by the fact that their peak and integrated flux density at 1.3\,cm are the same. These sources 
are not detected in the A configuration, even though the sensitivity is better. This implies that the source angular diameters are smaller
than the synthesized beam ($0\farcs24$) of the B configuration, but are partially resolved in the A configuration image. The latter
condition sets a lower limit to the angular diameters which are, respectively,  $0\farcs12$ (source 10),  $0\farcs11$ (source 11), $0\farcs11$ (source 12), and $0\farcs15$ (source 15).

Most of the radio sources are distributed in a N--S direction that roughly
coincides with that of the wall of the cavity of the NE--SW molecular outflow
mapped in CO\,(3--2) by  Gibb et al.~(\cite{gibb03}) and in  CO\,(7--6) by Qiu
et al.~(\cite{qiu13}). Based on the distribution of the radio sources, Gibb et
al.~(\cite{gibb03}) conclude that these sources trace a precessing thermal radio
jet and that such a radio jet is not driving the large-scale molecular outflow.
The sources likely associated with this N--S radio jet show an extended
morphology, with the exception of those located close to the geometrical center
(sources 6, 8, and 9), which have a more compact structure. One of these compact
sources, source 8 (resolved into 8a and 8b in the 1.3\,cm and 7\,mm maps), is associated with core B (Fig.~\ref{fig-alma}), the Keplerian
disk candidate associated with a B-type (proto)star identified by
S\'anchez-Monge et al.~(\cite{sanchez-monge13}).  As seen in Figs.~\ref{fig-cm}b
and c and \ref{fig-alma},  the centimeter emission of core B is resolved into
two sources at all wavelengths but 2\,cm.  Along the N--S structure, there is
one source (14) that has only been detected at 2\,cm (at a 9\,$\sigma$
level). Gibb et al.~(\cite{gibb03}) have detected a source at
6\,cm at $\alpha$(J2000)=18$^{\rm h}$ 58$^{\rm m}$ 13$\fs$07,
$\delta$(J2000)=+01$^{\circ}$ 40$'$ 40$\farcs$1 (source 4 in their notation) that is not visible at any other wavelength.
However, this source is detected at a 3\,$\sigma$ level, which casts some doubts
on the reliability of the detection.

Figure~\ref{fig-cm} also shows a group of three sources (1, 3, and 10) roughly 
aligned in the NE--SW direction. The morphology of these sources, in particular
sources 1 and 3, is very compact at all wavelengths.  These three sources could be
associated with the SiO high-velocity emission observed by Gibb et 
al.~(\cite{gibb03}) and  S\'anchez-Monge et al.~(\cite{sanchez-monge14}), which
could be tracing a second thermal radio jet. One of these
three radio sources (3) is associated with core A (Fig.~\ref{fig-alma}), the
other hot molecular core in \g35 that shows a velocity gradient suggestive of
rotation (S\'anchez-Monge et al.~\cite{sanchez-monge14}). This source is clearly
seen at 1.3\,cm and 7\,mm but only barely detected at 2\,cm. 

Finally, radio source 2 does not seem to be associated with
any of the (putative) outflows in the regions nor with dust continuum emission
(Fig.~\ref{fig-alma}). This source is very compact and visible at 2\,cm in the B
configuration and at 1.3\,cm and 7\,mm in the A configuration. The source has also been detected
at 6 and 3.6\,cm by Gibb et al.~(\cite{gibb03}) (see Table~\ref{table_cont2}).

\begin{table*}
\caption[] {Parameters of the Class I 44~GHz CH$_3$OH maser
features$^a$.}
\label{table_maser}
\begin{tabular}{ccccc}
\hline
&\multicolumn{2}{c}{Peak position}
\\
 \cline{2-3}  
&\multicolumn{1}{c}{$\alpha({\rm J2000})$} &
\multicolumn{1}{c}{$\delta({\rm J2000})$} &
\multicolumn{1}{c}{$I^{{\rm peak}, b}$} &
\multicolumn{1}{c}{$V_{\rm range}$} 
\\
\multicolumn{1}{c}{Feature} &
\multicolumn{1}{c}{h m s}&
\multicolumn{1}{c}{$\degr$ $\arcmin$ $\arcsec$} &
\multicolumn{1}{c}{(mJy/beam)} &
\multicolumn{1}{c}{(\kms)} 
\\
\hline
1       &18 58 11.580 &01 40 04.59 &177 &23--44 \\
2       &18 58 11.738 &01 40 32.08 &35  &23--44 \\
3       &18 58 11.744 &01 40 31.84 &58  &23--44 \\ 
4$^c$   &18 58 12.981 &01 40 33.74 &255 &23--44 \\
5       &18 58 13.011 &01 40 21.57 &47  &23--44 \\
6       &18 58 13.020 &01 40 20.31 &29  &30--44 \\
7       &18 58 13.053 &01 40 31.26 &41  &23--44 \\
8       &18 58 13.056 &01 40 28.96 &129 &23--44 \\
9       &18 58 13.063 &01 40 27.92 &20  &30--37 \\
10      &18 58 13.126 &01 40 30.97 &26  &23--44 \\ 
11      &18 58 13.127 &01 40 31.32 &19  &30--44 \\
12      &18 58 13.139 &01 40 26.92 &37  &23--44 \\
13      &18 58 13.347 &01 40 34.86 &47  &23--44 \\
14      &18 58 14.687 &01 40 32.01 &38  &23--44 \\
\hline
    
\end{tabular}

  $^a$ Positions and peak flux densities estimated from the B configuration map. \\
  $^b$ Peak intensity and integrated flux density corrected for primary beam
  response. \\
  $^c$ This feature is almost resolved into two spots in the A configuration map.
  The position and peak flux of the second spot are $\alpha({\rm J2000})$=18$^{\rm
  h}$ 58$^{\rm m}$ 12$\farcs$975,  $\delta({\rm J2000})$=01$\degr$ 40$\arcmin$
  33.80$\arcsec$, and 42\,mJy/beam.  \\
\\
\end{table*}

\subsection{CH$_3$OH masers at 44~GHz}

Class~I methanol masers are usually associated with high-mass star-forming
regions. Unlike the most common and bright Class~II methanol masers that are
radiatively pumped (e.g., Sobolev~\cite{sobolev07} and references therein), 
Class~I masers are collisionally pumped and  could be associated with
molecular outflows (Plambeck \& Menten~\cite{plambeck90}; Voronkov et
al.~\cite{voronkov06}). Our VLA  observations at 44~GHz have revealed the
presence of Class I CH$_3$OH maser emission in \g35. The presence of 44~GHz
CH$_3$OH maser emission in \g35 had been previously reported by Val'tts \&
Larionov~(\cite{valtts07}), but the positions of the maser spots had never been
published before.  A total of 14 maser spots have been detected in both
configurations, A and B. Their positions, peak intensities, and velocity ranges are
given in  Table~\ref{table_maser}. Unfortunately, the limited spectral
resolution of the observations ($\sim$7\,\kms) was not enough to separate
the different spectral components. In fact, except for one maser spot that has only been detected in one spectral channel and two that have been detected in two spectral channels, the rest have
been detected in three spectral channels covering a velocity range of 21~\kms. This suggests that most of the maser emission that
we observed is made of multiple velocity components. One of the spots, number 4, has an elongated morphology in the B configuration
and is barely resolved into two components in the A configuration.

\begin{figure*}
\centerline{
\hspace{4cm}
\includegraphics[angle=-90,width=20cm]{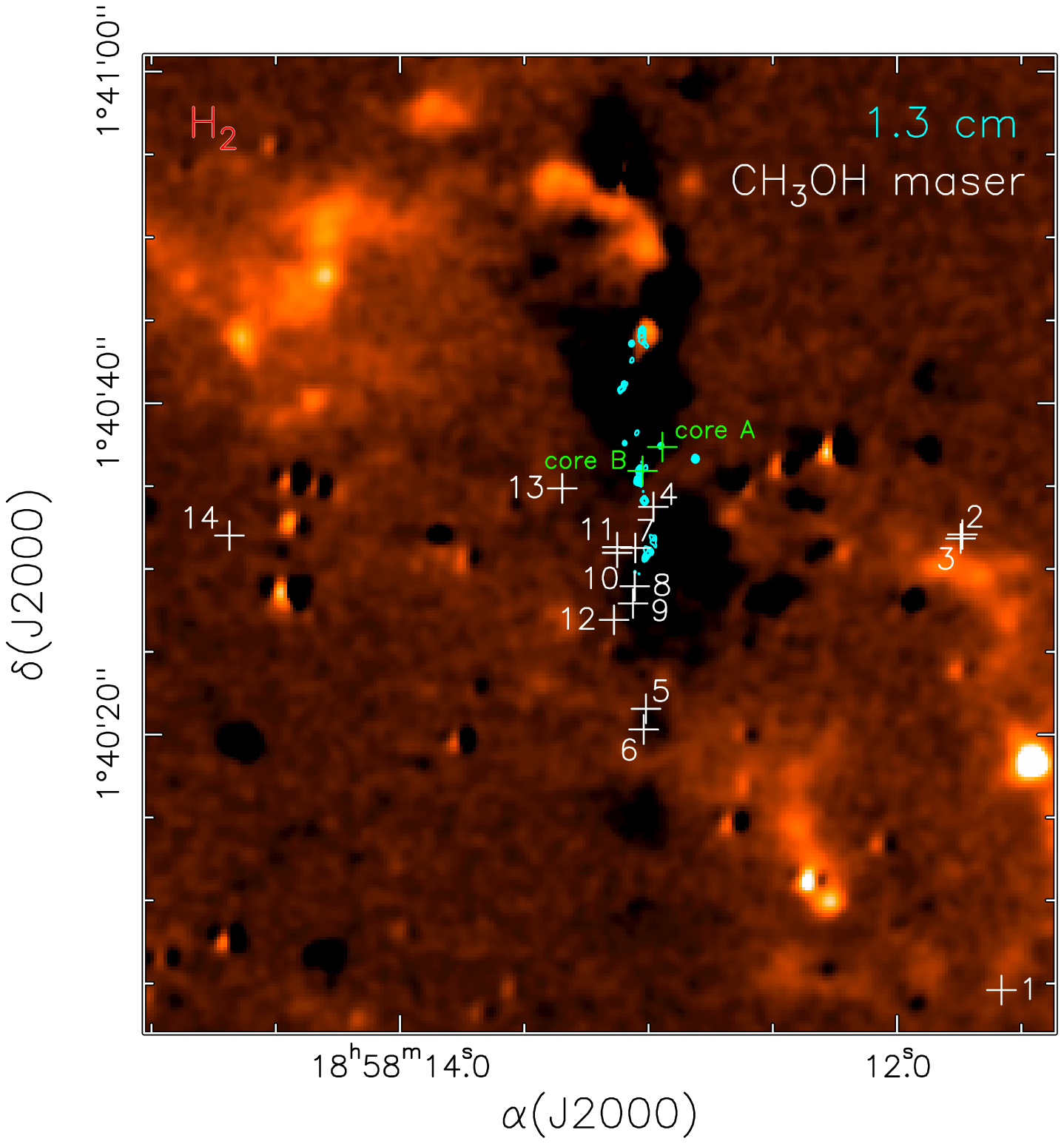}}
 \caption{H$_2$ 2.12\,$\mu$m emission towards \g35 from the UWISH2 survey (Froebrich et al.~\cite{froebrich11}) 
 overlaid by  the 1.3\,cm continuum emission ({\it cyan contours}) observed in the B configuration. 
Cyan contours are the same as in Fig.~\ref{fig-cm}. The numbers and white crosses indicate 
the positions of the Class I 44\,GHz CH$_3$OH masers (see
Table~\ref{table_maser}). Green crosses mark the positions of the 870\,$\mu$m continuum
sources core A and B (S\'anchez-Monge et al.~\cite{sanchez-monge14}).}
\label{fig-maser}
\end{figure*}

Figure~\ref{fig-maser} shows an image of the H$_2$ emission towards \g35  
overlaid by the 1.3\,cm continuum emission observed in the B configuration and the
positions of the 44\,GHz methanol masers. Almost all of the masers are located to
the south of core B and seem to follow the distribution of the centimeter
continuum sources and lie on the redshifted lobe of the CO outflow observed by
Gibb et al.~(\cite{gibb03}) and Qiu et al.~(\cite{qiu13}). This suggests that
the maser spots could trace the walls of the  cavity excavated by the outflow,
outlined by the H$_2$ emission (Fig.~\ref{fig-maser}). The only maser spots that
could be associated with the NE blueshifted lobe are numbers 13 and 
14.  There are also three maser spots, numbers 2, 3, and 14 located to the west and east of cores A and B. These masers could be 
associated with the walls of the lobes of the CO outflow. Alternatively, these maser spots 
could be tracing a second flow directed E--W and associated with the centimeter sources 1, 3, and 10, as suggested in the previous section.

\subsection{H$_2$O masers}
\label{h2o-masers}
\begin{figure}
\centerline{
\includegraphics[angle=0,width=9cm]{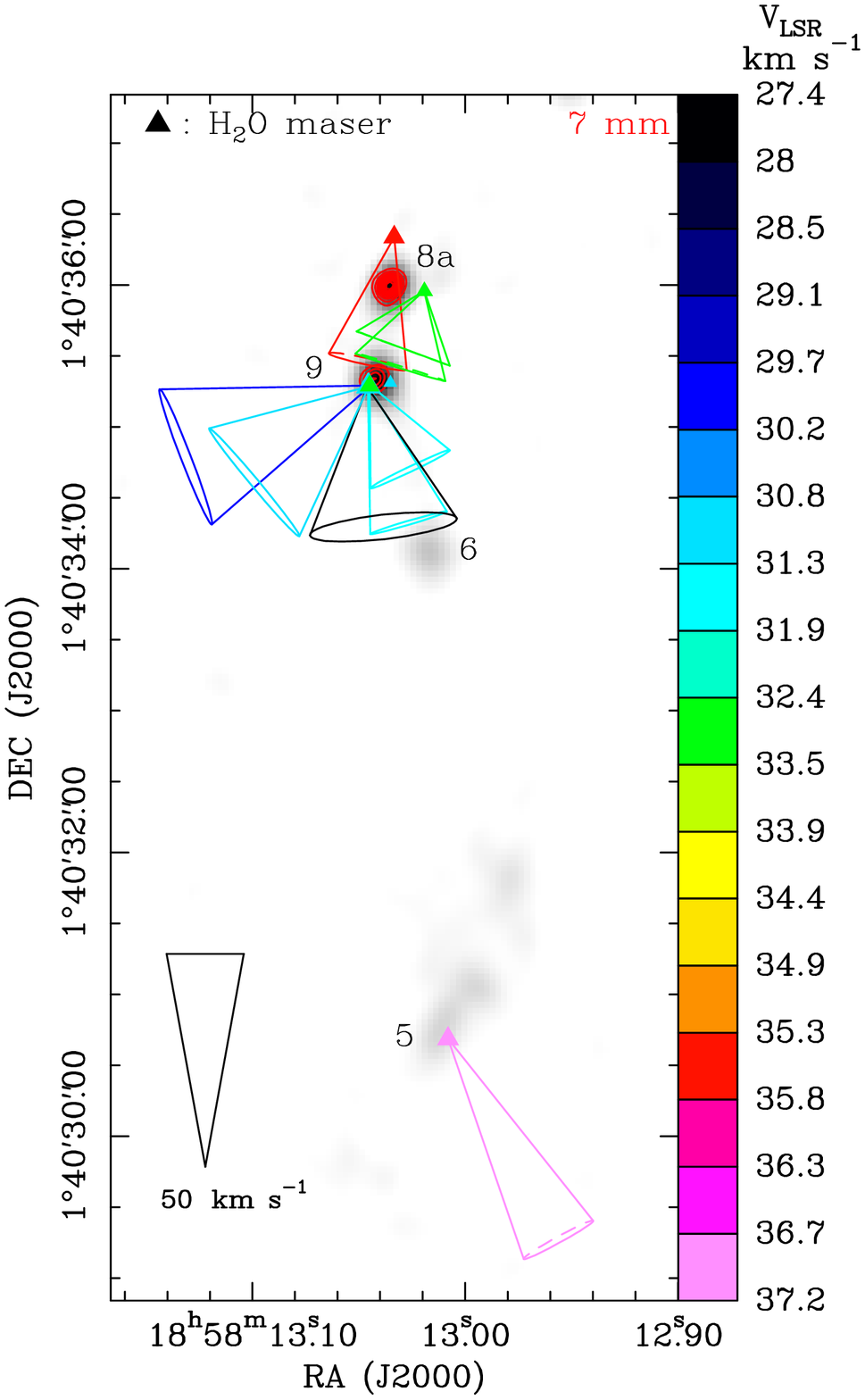}}
\caption{Distribution of water masers ({\it colored triangles}) towards \g35 overlaid with the
1.3\,cm ({\it grayscale image})  and 7\,mm ({\it red contours}) VLA continuum observed in the B configuration. 
Colored triangles show the absolute position of individual maser features, 
with colors denoting the maser $V_{\rm LSR}$ according to the color-velocity conversion code
shown on the right side of the panel. 
The triangle area is proportional to the logarithm of the maser
intensity.  
 {\it Colored cones} represent the measured maser proper motions,
with cone aperture giving the uncertainty on the direction of the motion.
 The amplitude scale for the maser velocity is indicated by the {\it black cone} at the
bottom left of the panel.
The grayscale image reproduces linearly the intensity of the VLA B-Array 1.3\,cm continuum
from \ 0.05~mJy~beam$^{-1}$ ($\approx$ 3~$\sigma$) to \ 1.7~mJy~beam$^{-1}$.
The red contours reproduce the  B-Array 7\,mm continuum, plotting levels 
from  \ 10\% to 90\%, in steps of 10\%, of the peak value of \ 3.5~mJy~beam$^{-1}$.
Radio knots are labeled using the same numbers as in Table~\ref{table_cont1}.
}
\label{fig-water}
\end{figure}

Water masers are typically associated with winds/jets from both low-~and~high-mass YSOs.
They are thought to be collisionally excited in relatively slow ($\le30$~\kms) C-shocks and/or fast ($\ge30$~\kms) J-shocks (Gray~\cite{gray12}; Hollenbach et al.~\cite{hollenbach2013}) produced by the interaction of the protostellar outflow with very dense (n$_{H_2}$ $\ge$ 10$^7$~\cmc) circumstellar material.
Water maser emission towards \g35 was previously observed with the VLA (see, e.g., Codella et al.~\cite{codella10}), but to our knowledge
no Very Long Baseline Interferometry (VLBI) observations of the 22~GHz masers towards this source have been reported in the literature.
Figure~\ref{fig-water} reports absolute positions and velocities of the water maser features detected with our VERA observations. 
For a description of the criteria used to identify individual masing clouds, to derive their
parameters (position, intensity, flux, and size), and to measure their proper motions
(relative and absolute), see the paper
on VLBI observations of water and methanol masers by Sanna et al.~(\cite{sanna10}). 
The derived absolute water maser proper motions were corrected for the
apparent proper motion due to the Earth's orbit around the Sun
(parallax), the solar motion, and the differential Galactic rotation
between our LSR and that of the maser source. We have adopted a flat
Galaxy rotation curve ($R_0 = 8.33\pm0.16$~kpc, $\Theta_0 = 243\pm6
$~\kms) (Reid et al.~\cite{reid14}) and the solar motion ($U = 11.1^{+0.69}_{-0.75}$, 
$V = 12.24^{+0.47}_{-0.47}$, and $W = 7.25^{+0.37}_{-0.36}$~\kms) by Sch{\"o}nrich et al.~(\cite{sch10}), who recently revised 
the Hipparcos satellite results.

Table~\ref{tab-water} reports the parameters of the twelve detected maser features.
Water maser emission extends N--S across \ $\approx$5\farcs6 with a close spatial correspondence with the central portion of the radio jet. 
In fact, most of the maser features are observed
near the radio knots 8a, 9, and 5. Maser proper motions are mainly pointing to the south with amplitudes in the range \ 10--50~\kms,
that apparently increase from north to south.

\begin{table*}
\caption{\label{tab-water} 22.2~GHz H$_2$O maser parameters for G35.20$-$0.74N}
\centering                          
\begin{tabular}{ccccrrrr}       
\hline             
\multicolumn{1}{c}{Feature} 
& \multicolumn{1}{c}{Epochs of} 
& \multicolumn{1}{c}{I$_{\rm peak}$} 
& \multicolumn{1}{c}{$V_{\rm LSR}$} 
& \multicolumn{1}{c}{$\Delta~x$} 
& \multicolumn{1}{c}{$\Delta~y$} 
& \multicolumn{1}{c}{$V_{x}$} 
& \multicolumn{1}{c}{$V_{y}$}\\
\multicolumn{1}{c}{Number}  
& \multicolumn{1}{c}{Detection}
 & \multicolumn{1}{c}{(Jy beam$^{-1}$)} 
 & \multicolumn{1}{c}{(km s$^{-1}$)} 
 & \multicolumn{1}{c}{(mas)} 
 & \multicolumn{1}{c}{(mas)} 
 & \multicolumn{1}{c}{(km s$^{-1}$)} 
 & \multicolumn{1}{c}{(km s$^{-1}$)}\\
\hline 
    1 &          2,3,4 &     80.48 &    31.0 &   179.11$\pm$0.07 & $-$1052.40$\pm$0.08 &   26.9$\pm$11.3 &  $-$22.5$\pm$11.3  \\
    2 &        1,2,3,4 &     63.56 &    31.4 &   184.32$\pm$0.07 & $-$1051.20$\pm$0.07 &   $-$9.5$\pm$ 6.6 &  $-$32.1$\pm$ 6.6  \\
    3 &        1,2,3,4 &     31.22 &    37.2 &  $-$379.16$\pm$0.07 & $-$5650.49$\pm$0.07 &  $-$25.9$\pm$ 6.6 &  $-$47.2$\pm$ 6.6  \\
    4 &        1,2,3,4 &     27.33 &    35.5 &     0.00$\pm$0.00  &     0.00$\pm$0.00   &    6.2$\pm$ 6.6  &  $-$29.2$\pm$ 6.6  \\
    5 &              4 &     15.07 &    32.5 &   173.36$\pm$0.07 & $-$1047.34$\pm$0.07 &      ... $ \; \; \; \; \, $ &      ... $ \; \; \; \; \, $  \\
    6 &        1,2,3,4 &      7.20 &    31.7 &   172.58$\pm$0.08 & $-$1041.02$\pm$0.08 &   $-$9.5$\pm$ 6.6 &  $-$19.8$\pm$ 6.7  \\
    7 &        1,2,3,4 &      6.84 &    33.0 &  $-$213.52$\pm$0.07 &  $-$380.36$\pm$0.07 &    5.0$\pm$ 6.6 &  $-$13.5$\pm$ 6.6  \\
    8 &          2,3,4 &      5.77 &    27.4 &   188.74$\pm$0.07 & $-$1055.96$\pm$0.07 &   $-$3.8$\pm$11.2 &  $-$32.9$\pm$11.2  \\
    9 &            3,4 &      5.10 &    31.7 &   185.80$\pm$0.07 & $-$1054.28$\pm$0.07 &      ... $ \; \; \; \; \, $ &      ... $ \; \; \; \; \, $  \\
   10 &          2,3,4 &      4.36 &    29.8 &   164.41$\pm$0.08 & $-$1045.77$\pm$0.08 &   43.5$\pm$11.3 &  $-$16.8$\pm$11.4  \\
   11 &              4 &      3.39 &    30.9 &    25.88$\pm$0.08 & $-$1026.68$\pm$0.09 &      ... $ \; \; \; \; \, $ &      ... $ \; \; \; \; \, $  \\
   12 &        1,2,3,4 &      1.63 &    33.4 &  $-$210.77$\pm$0.08 &  $-$377.03$\pm$0.08 &    5.6$\pm$ 6.7 &  $-$18.0$\pm$ 6.7  \\
\hline
\end{tabular}
\tablefoot{\\
Column~1 gives the feature label number; Col.~2
lists the observing epochs at which the feature was detected;
Cols.~3~and~4 provide the intensity of the strongest spot
and the intensity-weighted LSR velocity, respectively, averaged over the
observing epochs; Cols.~5~and~6 give the position offsets (with
the associated errors) along the R.A. and Dec. axes, relative to the
feature 4, measured at the first epoch of detection; Cols.~7~and~8 give the components of the absolute
proper motion (with the associated errors) along the R.A. and Dec. axes. \\
The absolute position of the feature 4 at the epoch on  September 6, 2013, is: 
$\alpha$(J2000) = 18$^{\rm h}$ 58$^{\rm m}$ 13\fs0334, $\delta$(J2000) = 1$\degr$ 40$^{\prime}$ 36\farcs337,
with an accuracy of \ $\pm$2~mas. 
}
\end{table*}

\section{Analysis}

\subsection{Continuum spectra}
\label{spectral}

\begin{figure*}
\centerline{
\includegraphics[angle=-90,width=18cm]{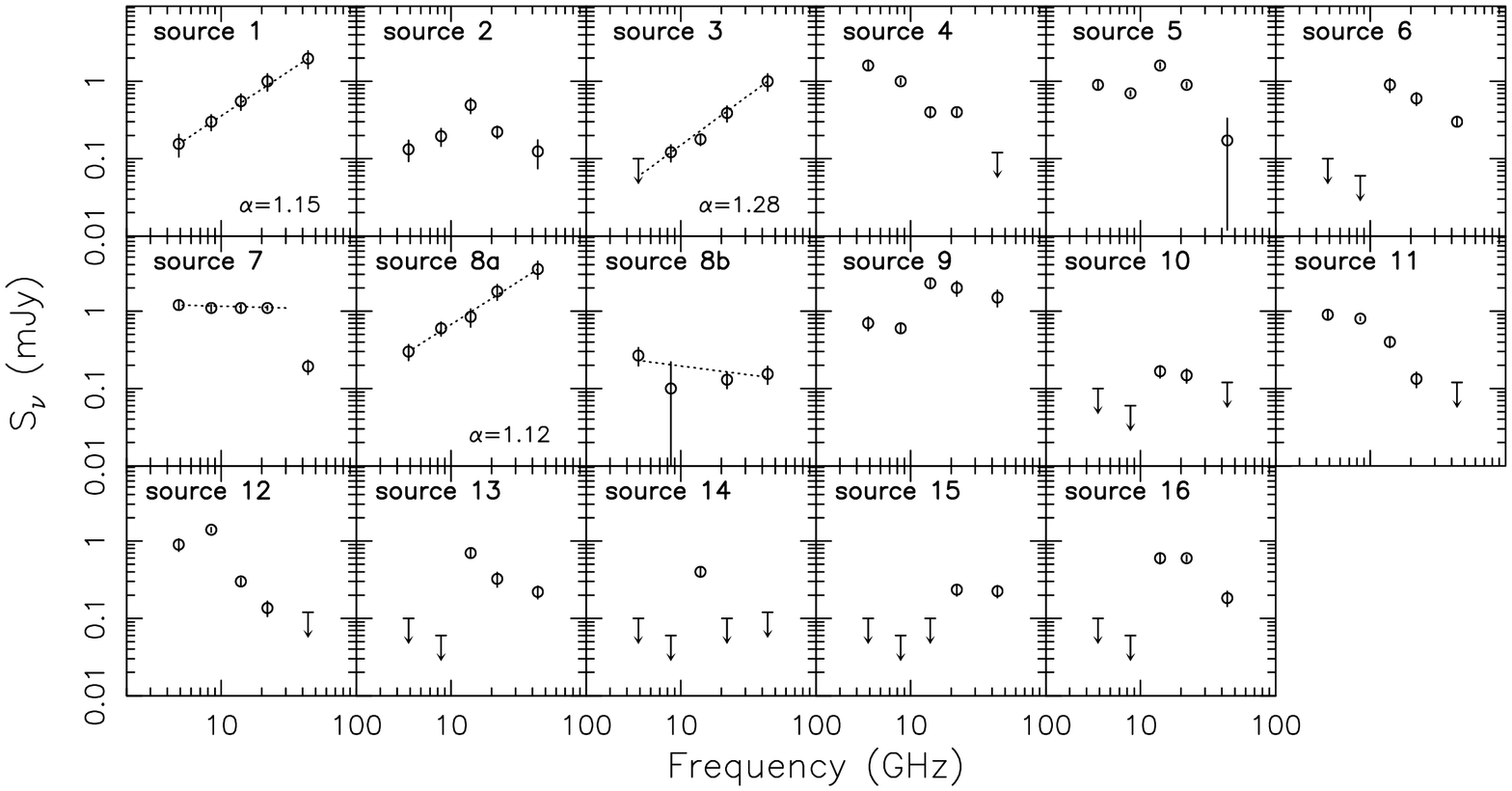}}
\vspace*{-1.5cm}
\caption{Continuum spectra for the centimeter continuum sources.  
Circles and upper limits correspond to the observational data from
Tables~\ref{table_cont1} and \ref{table_cont2}. For sources 1, 3, and 8a the
dotted line indicates the linear fit ($S_\nu\propto\nu^\alpha$) from 6\,cm to
7\,mm (for source 3 the fit was done from 3.6\,cm to
7\,mm).}
\label{fig-index}
\end{figure*}

Figure~\ref{fig-index} shows the continuum spectra for all the
centimeter sources detected in at least one wavelength. The spectra have been
estimated with the fluxes tabulated in Tables~\ref{table_cont1} and
\ref{table_cont2}. For most of the sources the emission is so compact at all
wavelengths and angular resolutions that the integrated fluxes hardly change when the 
maps are re-constructed with the same uvrange and restoring beam. As seen in
this figure, only three sources (1, 3, and 8a) have spectral index $\alpha>1$
(assuming $S_\nu\propto\nu^\alpha$). For source 7, without taking into
account the emission at 7\,mm, $\alpha\simeq0$ is consistent with optically thin
free-free ($\alpha=-0.1$) emission. This source has a flux density at 7\,mm of
0.19\,mJy in the B configuration, but it is not detected at a 3\,$\sigma$ level
of 0.10\,mJy in the A configuration.  
A possible explanation for such a faint emission at 7\,mm, as compared to the
emission at the other wavelengths, could be that the source is very extended and
its emission has  almost been resolved out by the interferometer. This is
supported by the fact that the source is barely detected at 1.3\,cm and is not
detected at 7\,mm at the higher angular resolution  provided by the A
configuration. Source 8b has a spectral index $\alpha\simeq-0.2$, which is
also consistent with optically thin free-free emission. The emission of
this source at some wavelengths is very weak and very difficult to
separate from the stronger emission of 8a. 
In fact, at 2\,cm it has not
been possible to estimate it. Therefore, the flux density estimates are  more
uncertain.

For most of the sources, the emission shows a negative spectral index for
wavelengths shorter than 3.6 or 2\,cm. For some of these, the continuum spectra shows a double
slope, sharply  increasing at longer wavelengths and then decreasing for shorter
ones (e.g., source 2). A possible explanation for this behavior could be that the
observations at 6 and 3.6\,cm, carried out  at a very different epoch (1999)
than ours (2013--2014), had some calibration problems. However, this seems inconsistent with
the fact that for some sources (e.g., 1 or 8a) the 6 and 3.6\,cm
fluxes are in agreement with the extrapolation of the fluxes at shorter
wavelengths. Another possibility could be variability of the sources. Since
the observations at longer and shorter wavelengths were 
separated by almost 15 years, variability could affect the estimated fluxes and the
spectral index determinations. In fact, as suggested in Sect.~\ref{cont-em}, source 2 could be variable.
A strong variability is characteristic of
extragalactic sources. Following Anglada et al.~(\cite{anglada98}), the expected
number of background extragalactic sources at 1.3\,cm within a field of diameter
2$'$ (the primary beam at 1.3\,cm) is given by $<N>=0.013\,S_0^{-0.75}$, where
$S_0$ is the detectable flux density threshold at the center of the field
in mJy. For $S_0$=0.10\,mJy (3\,$\sigma$ in the B configuration), the expected
number of background extragalactic sources is 0.07. A similar value is obtained
at 6\,cm using the same field and the same $S_0$, which is also 3\,$\sigma$
at 6\,cm. Only using a field at 6\,cm with a diameter $>4\farcs5$, which is much
larger than the field in which the radio sources have been detected, the
expected number of background sources would be one.   

This result is in agreement with the fact that most of the centimeter sources
should be associated with the thermal radio jet. In principle, thermal sources
should not have strong variability, so this could indicate that the emission of
some of the knots of the radio jet is non-thermal. In particular, Rodr\'{\i}guez
et al.~(\cite{rodriguez89}) measured a negative spectral index of $\alpha=-0.7$
for the radio jet powered by the YSO S68 FIRS1, which has been interpreted as
produced by non-thermal emission. More recently, Carrasco-Gonz\'alez et
al.~(\cite{carrasco-gonzalez10}) have revealed synchrotron emission arising from
the HH\,80--81 jet.  In fact, the presence of synchrotron emission in radio jets might be more common
 than originally expected, as has been shown by the recent results of Moscadelli et al.~(\cite{moscadelli13}, \cite{moscadelli16}).
Therefore, non-thermal variable emission could be a plausible
explanation for the negative spectral indexes and the strange continuum spectra obtained for
most of the sources. In order to test the variability of the sources, new
observations at 6 and 3.6\,cm are needed.  

Source 14 is the only one that has been detected at only one wavelength
(2\,cm) at a 9\,$\sigma$ level. 
A possible explanation for the non-detection at 1.3\,cm and 7\,mm is that
the source, which is slightly extended at 2\,cm, has been resolved out by the
interferometer at these wavelengths. 
Regarding the non-detection at 6 and 3.6\,cm, the expected peak flux densities, estimated by 
extrapolating the peak flux density measured at 2\,cm (0.28\,\mjy) with a spectral index of 1 and taking 
into account the size of the source and the different synthesized beams, are 0.11 and 0.05\,\mjy, respectively.  These fluxes are similar to or below
the 3\,$\sigma$ level and are therefore consistent with the fact that source 14 has not been detected at these wavelengths.

In conclusion, most of the radio sources in \g35 have spectral indices that
are either negative (i.e., the flux decreases with frequency)  or consistent with source
variability. In both cases, the most likely explanation is that their emission
is non-thermal. The only sources that have spectral indices consistent with
free-free thermal emission from ionized gas are sources 3, 8a, and 8b associated
with the hot molecular cores A and B (Figs.~\ref{fig-alma} and \ref{fig-zoom}), and sources 1 and 7.

\section{Discussion}

The main goal of our high-angular resolution centimeter observations was to
investigate the existence of a binary system at the center of the massive core B
in \g35. Such a system could explain the mass of 18\,$M_\odot$, computed from the velocity
field, and the possible precession of the radio jet. The latter
is supported by the S-shaped morphology of the jet and by the jet position angle, which is
very different from that of the CO outflow (e.g., Heaton \&
Little~\cite{heaton88}; Gibb et al.~\cite{gibb03}). 
Another feature of the radio 
jet is that its expansion, suggested by the proper motions of the  H$_2$O masers  (see Sect.~\ref{h2o-masers}), can also be estimated  from the centimeter continuum emission 
at different epochs. All these issues will be discussed in detail in Sects.~\ref{binary} to 
\ref{precession}. In the next section we investigate the nature of the free-free emission of 
source 8a, associated with core B, and source 3, associated with core
A. We also discuss the nature of source 1, which could be powering an additional jet in the region.

\subsection{Origin of the free-free ionized gas emission}
\label{nature}

The only three sources that have a positive spectral index consistent with
thermal emission from ionized gas are 1, 3, and 8a. Their spectral index is
1.1--1.3, consistent with partially optically thick emission
arising from a thermal radio jet (e.g., Anglada~\cite{anglada96}) or an
ultracompact or hypercompact \HII (\UCHC) region with a density gradient (e.g., Franco et
al.~\cite{franco90}). In the following we discuss these two possibilities and
explore the properties of the sources for each scenario. 

\subsubsection{Radio jet}

Source 8a is located at the center of the N--S radio jet, first mapped by
Gibb et al.~(\cite{gibb03}), while sources 3 and 1 lie at the center and edge,
respectively, of what could be a second radio jet centered on core A (see S\'anchez-Monge et al.~\cite{sanchez-monge14}).
Typical spectral indices of radio jets are $\alpha\lesssim$1.3
(Anglada~\cite{anglada96}). Therefore, the emission of the three sources could
be in fact arising from free-free thermal emission of radio jets. In this
scenario the gas is ionized by the UV photons emitted by the shock at the
interface between a neutral stellar wind and the surrounding high-density gas (Torrelles et
al.~\cite{torrelles85}). 

Reynolds~(\cite{reynolds86}) modeled the emission of collimated, ionized stellar
winds. According to this model, for a radio jet with constant velocity, temperature, and ionization
fraction, the spectral index of the emission is related
to its collimation  through the expression $\alpha=1.3-0.7/\epsilon$, where
$\epsilon$ is the power-law
index of the dependence of the jet radius on the distance from the central
object. When the emission
of the three sources arises from radio jets, its collimation $\epsilon$ would be $>$3.9
(Table~\ref{table_jet}). Therefore, these radio sources would have an increasing
opening angle with distance from the star and would be poorly confined at the small scales
of a few hundreds of au at which the centimeter observations are sensitive. The
model of Reynolds~(\cite{reynolds86}) also estimates the injection
radius $r_0$, i.e., the distance at which the ionized flow originates,
assuming an isothermal radio jet. This distance determines the turnover
frequency $\nu_m$ above which the entire jet becomes optically thin. Following
Eq.~(1) of Beltr\'an et al.~(\cite{beltran01}), we estimated $r_0$ from the
1.3\,cm B configuration data  assuming an electron temperature $T=10^4$\,K and  
$\theta_0$=1\,rad and $\sin i=1$, where $\theta_0$ is
the jet injection opening angle and $i$  is the jet axis inclination with
respect to the line of sight. 
The assumption  $\sin i=1$ implies that the jet lies on the
plane of the sky and in some particular cases this can
produce a significant underestimate of the injection
radius $r_0$. However, for a random orientation of the jet in space, the
average value of $\sin i$ is $\pi/4$, which changes the $r_0$
estimate only by 13\%. The assumption $\theta_0$=1\,rad implies that the jet is
poorly collimated (according to Reynolds~\cite{reynolds86}, well-collimated jets 
have  $\theta_0\lesssim0.5$), in agreement with what is suggested by the
collimation parameter $\epsilon$. Figure~\ref{fig-index} shows that there is 
no hint of turnover in the spectra of sources 8a, 3, and 1. Therefore, we assumed  
43.94~GHz ($\lambda=$7\,mm) to be a lower limit to the turnover frequency
and estimated an upper limit for the injection radius. The values of $r_0$
obtained are 50--90~au (Table~\ref{table_jet}), indicating that the ionization
of the jet begins at a small distance from the star. 

The model of Reynolds~(\cite{reynolds86}) also allows us to estimate the ionized
mass-loss rate in the jet $\dot M_{\rm ion}$. Following Eq.~(3) of Beltr\'an et
al.~(\cite{beltran01}), we estimated $\dot M_{\rm ion}$ for a pure hydrogen jet,
assuming a stellar wind velocity $V_\star$ of 200\,\kms, $\theta_0$=1\,rad, and
$\sin i=1$. For a random orientation of the jet in space, the mass-loss rate in the jet would be
underestimated only by 6\%. The values of $\dot M_{\rm ion}$ are
1.4--3.2$\times10^{-6}$\,$M_\odot$\,yr$^{-1}$ (Table~\ref{table_jet}). 
These values are 
consistent with the values of
$10^{-6}$\,$M_\odot$\,yr$^{-1}$ found by Ceccarelli et al.~(\cite{ceccarelli97})
in low-mass Class\,0 objects and are 2--3 orders of
magnitude higher than those found by  Beltr\'an et al.~(\cite{beltran01}) in
more evolved low-mass young stellar objects.  We note that these values have to be taken as
lower limits, not only because the turnover frequency is a lower limit, but
also because the stellar wind velocity could be
higher for high-mass YSOs.

\begin{table*}
\caption[] {Parameters of the radio jets$^a$.}
\label{table_jet}
\begin{tabular}{ccccc}
\hline
& & &
\multicolumn{1}{c}{$r_0$} &
 \multicolumn{1}{c}{$\dot M_{\rm ion}$} 
\\
\multicolumn{1}{c}{Source} &
\multicolumn{1}{c}{$\alpha$}&
\multicolumn{1}{c}{$\epsilon$} &
\multicolumn{1}{c}{(au)} &
\multicolumn{1}{c}{($M_\odot$\,yr$^{-1}$)} 
\\
\hline
1       &1.15 &4.7  &$<$ 52 &$>$ 1.4$\times10^{-6}$\\
3       &1.28 &35   &$<$ 89 &$>$ 3.2$\times10^{-6}$\\
8a      &1.12 &3.9  &$<$ 63 &$>$ 1.9$\times10^{-6}$\\ 
\hline
    
\end{tabular}

  $^a$ In the case the free-free emission of these sources arises from shock
induced ionization. \\
\\
\end{table*}

\subsubsection{\UCHC\ region}

\begin{figure}
\centerline{
\hspace{.5cm}
\includegraphics[angle=0,width=11.5cm]{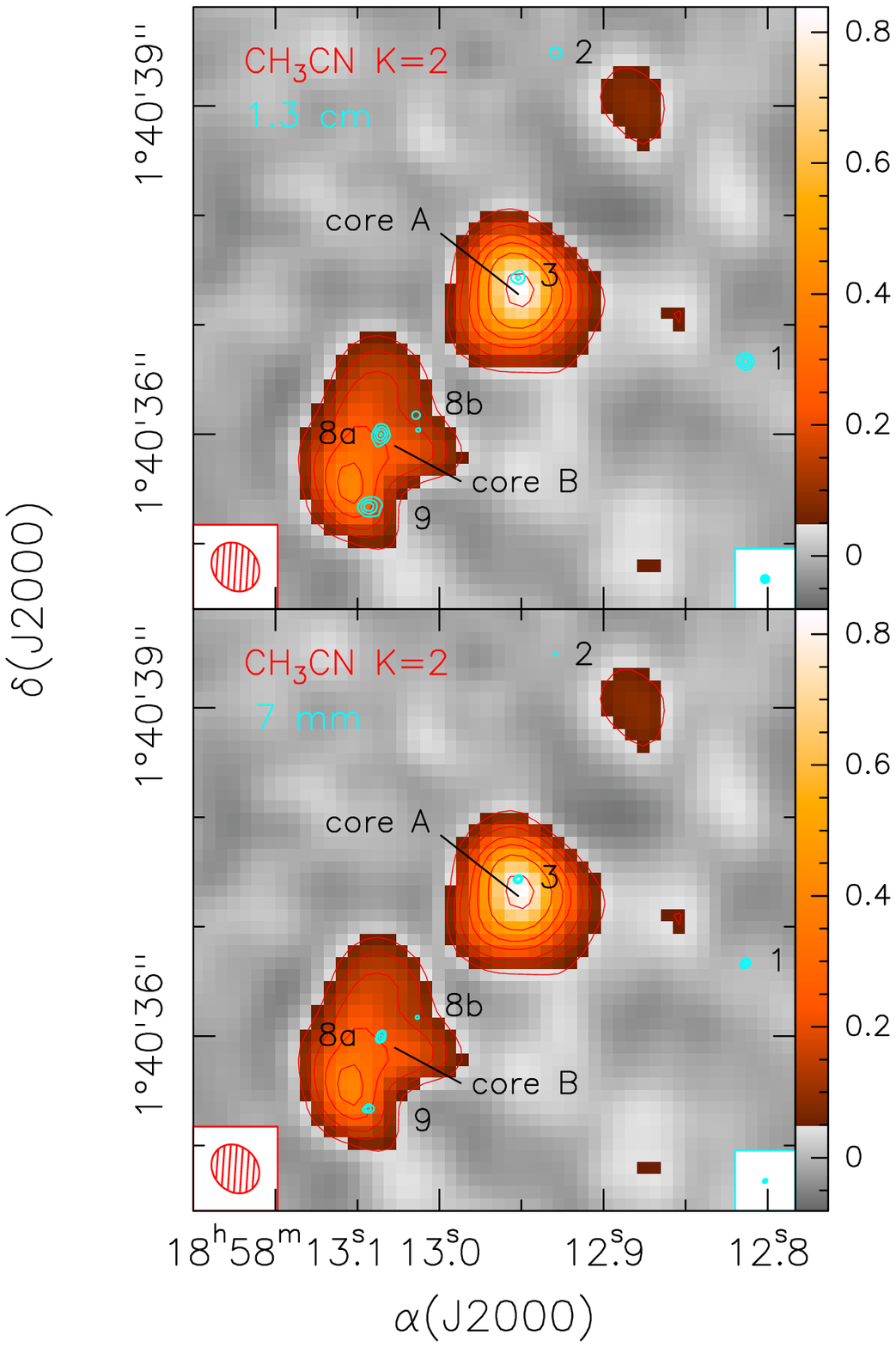}}
\vspace{-1cm}
\caption{({\it Top panel}) Overlay of the 1.3\,cm continuum emission ({\it cyan contours}) 
observed in A configuration on the average emission in the CH$_3$CN\,(19--18) K=2
line ({\it colors} and {\it red contours}) at 350~GHz observed with ALMA by S\'anchez-Monge et
al.~(\cite{sanchez-monge13}). The cyan contours are 3, 12, 30, and 54 
times 1\,$\sigma$, which is 0.023\mjy. The velocity interval used to calculate the
CH$_3$CN mean emission is 25.9--33.0\,\kms. The red contour levels are 1, 2, 3, 5, 7, 10, and
15 times 50\mjy. The
ALMA and VLA synthesized beams are shown in the lower left and lower right 
corner, respectively. The numbers correspond to the centimeter continuum sources in 
Table~\ref{table_cont2}. ({\it Bottom panel}) Same for the  7\,mm continuum
emission. The cyan contours are 3, 12, and 48 
times 1\,$\sigma$, which is 0.033\mjy.} 
\label{fig-zoom} 
\end{figure}

The free-free emission of sources 1, 3, and 8a could also be
explained in terms of photoionization.  
Thermal radio jets are expected to have an extended and elongated
morphology at subarcsecond angular resolution (e.g.,
Anglada~\cite{anglada96}) and lower flux at higher angular resolution because part of the extended emission is filtered out by the interferometer.
In fact, the typical sizes (a few 100\,au) of
radio jets associated with high-mass protostars (e.g., Anglada et al.~\cite{anglada14}) should be easily resolved
at the angular resolution of $\sim$$0\farcs05$ ($\sim$100\,au at the distance of \g35) reached at 7\,mm in the A configuration. 
In contrast, the observations of Gibb et al.~(\cite{gibb03}), with a maximum angular resolution of $\sim$$0\farcs25$ at 3.6\,cm,
were not able to properly resolve the emission and distinguish between a radio jet or an \UCHC\ region origin.  As seen in Fig.~\ref{fig-zoom}, none of the
three free-free sources is resolved. On the contrary, the sources 
are very compact  and their fluxes do not change from the B to the A configuration.
 Furthermore, two of the free-free sources, 3 and 8a, are
associated respectively with the hot molecular cores A and B (see
Fig.~\ref{fig-zoom}). This also favors the interpretation of the emission as being due to photoionization because hot cores are often known to be associated with 
\UCHC\ regions.

Assuming that the centimeter continuum emission comes from homogeneous optically
thin \UCHC\ regions, we used the data at 1.3\,cm obtained with the B configuration to calculate
the physical parameters of the three sources
(using the equations of Mezger \& Henderson~\cite{mezger67} and
Rubin~\cite{rubin68}; see also the Appendix of Schmiedeke et al.~\cite{schmiedke16}). 
Table~\ref{table_uc} gives the spatial radius $R$ of the
\UCHC\ region obtained from the deconvolved source size,  the source
averaged brightness $T_B$, the  electron density $n_e$, the emission measure
$EM$, the number of Lyman-continuum photons per second $N_{\rm Ly}$, the  mass
of ionized gas $M_{\rm ion}$, calculated assuming a spherical
homogeneous distribution, and the spectral type of the ionizing star. The
last was computed from the estimated $N_{\rm Ly}$ using the tables
of Davies et al.~(\cite{davies11}) and Mottram et al.~(\cite{mottram11}) for  
zero age main sequence (ZAMS) stars. 

As seen in Table~\ref{table_uc}, sources 1, 3,
and 8a are ionized by stars of spectral type
B1--B2, with masses of 8--11\,$M_\odot$ (Mottram et
al.~\cite{mottram11}). However, it has to be taken into account that the parameters
of the \UCHC\ regions have been estimated assuming optically thin emission while
the spectral indices $>1.1$ indicate that the emission is partially optically
thick. In this case, $N_{\rm Ly}$, and therefore, the spectral type and the mass of the central objects, 
should be considered lower limits. 

If source 8a is indeed an \UCHC\ region and at the same time is the
driving source of the radio jet (as suggested by its location at
the geometrical center of the jet),  then G35.20$-$0.74N would represent a
unique example of a radio jet coexisting with an \UCHC\ region powered by the same
YSO. This suggests that this object is in a transitional stage between the main
accretion phase, dominated by infall and ejection of material, and the phase in
which the radiation pressure of the ionized circumstellar material overcomes infall and the
newly formed \UCHC\ region begins to expand (see Keto~\cite{keto02}).  Guzm\'an et al.~(\cite{guzman16})
have very recently reported a similar case for G345.4938+01.4677. A similar scenario could be also
valid for source 3, but in this case the existence of a radio jet is
less clear. In this scenario, one of the knots of the jet would be source 1, 
but as discussed above,  this centimeter source could also be an \UCHC\ region;
in fact, Fuller et al.~(\cite{fuller01}) propose source 1 as the driving source of
an additional jet in the region, observed at near-infrared wavelengths. However, the association of  source 3 with a dense core with signs of
rotation would favor it as the driving source of the possible jet.

Based on the compact morphology of sources 1, 3, and 8a, which are not resolved even at a spatial resolution of $\sim$100\,au, 
and the poor 
collimation of the emission, we conclude that 
the most likely origin of their free-free emission is photoionization. Therefore, we believe that the three 
sources are likely associated with \UCHC\ regions.

\begin{figure}
\centerline{
\includegraphics[angle=-90,width=9cm]{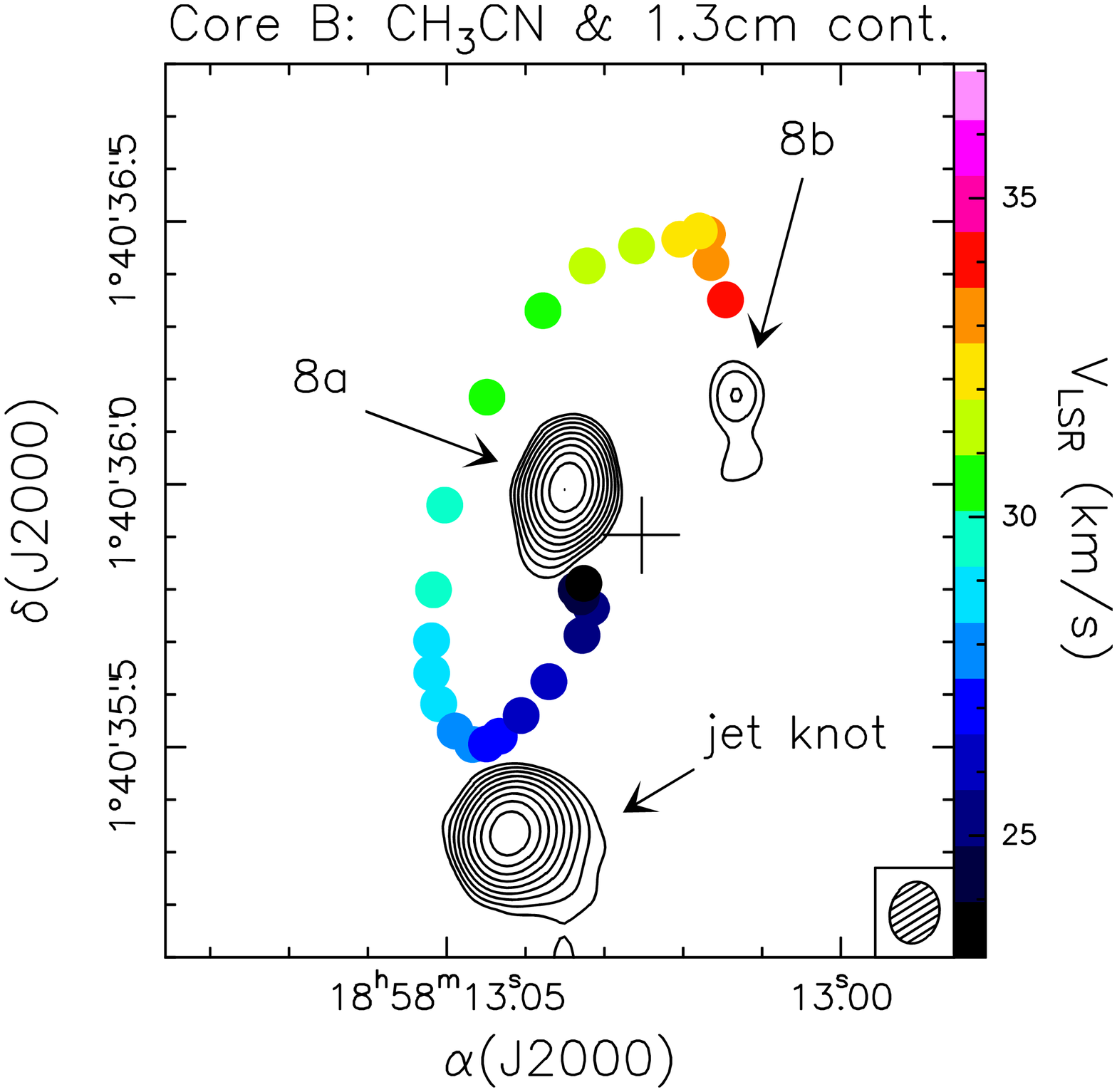}}
\caption{VLA map of the 1.3~cm continuum emission (contours) towards \g35 overlaid on the 
Keplerian disk traced by the peaks of the CH$_3$CN(19--18) $K$=2 emission at different velocities 
(solid circles, color-coded as indicated in the figure), imaged with ALMA (see S\'anchez-Monge et al.~2013). 
The cross marks the position of the sub-mm continuum emission at 350~GHz. Contour levels range in 
logarithmic steps from 0.06 to 1.452~mJy/beam. The synthesized beam of the VLA is shown at the bottom right. } 
\label{fig-disk} 
\end{figure}

\begin{table*}
\caption[] {Physical parameters of the \UCHC\ regions$^a$.} 
\label{table_uc}
\begin{tabular}{ccccccccc}
\hline
&\multicolumn{1}{c}{$R$} &
\multicolumn{1}{c}{$T_B$} &
\multicolumn{1}{c}{$n_e$} &
\multicolumn{1}{c}{$EM$} &
\multicolumn{1}{c}{$N_{\rm Ly}$} & 
\multicolumn{1}{c}{$M_{\rm ion}$} &
\multicolumn{1}{c}{Spectral} 
\\
\multicolumn{1}{c}{Source} &
\multicolumn{1}{c}{($10^{-3}$ pc)}&
\multicolumn{1}{c}{(K)}&
\multicolumn{1}{c}{($10^5$ cm$^{-3}$)} &
\multicolumn{1}{c}{($10^6$ cm$^{-6}$ pc)} &
\multicolumn{1}{c}{($10^{45}$ s$^{-1}$)} &
\multicolumn{1}{c}{($10^{-6}$ $M_\odot$)}&
\multicolumn{1}{c}{Type} 
\\  
\hline
 1                    & 0.2    &1440    &$10$	  &329   &0.6 &1.2    & B2 \\
 3$^b$            &$<$1.3  &$>$17   &$>0.5$  &$>4$  &0.2 &$<$10  & B2   \\
 8a                  & 0.3    &1170    &$7.8$   &266   &1.0 &2.9    & B1   \\ 
 8b$^b$         &$<$1.3    &$>$6   &$>0.3$   &$>1.3$   &0.08 &$<$6    & B3   \\ 
\hline
\end{tabular}
\\
$^a$ In the case the free-free emission of these sources arises from 
photoionization. Estimated from the 1.3\,cm data in B configuration. \\
$^b$ Unresolved source. The upper limit corresponds to the beam size at 1.3\,cm
in B configuration. 
\end{table*}

\subsection{Binary system in core B}
\label{binary}

S\'anchez-Monge et al.~(\cite{sanchez-monge13}) observed \g35 with ALMA at
870\,$\mu$m and discovered a Keplerian rotating disk associated with core B. The
best-fit model to the velocity field gives a mass of the central star of
18\,$M_\odot$. However, this poses a problem because the luminosity expected of
such a massive star would be comparable to (or higher than, depending on the
models used to estimate it) the luminosity of the whole star-forming region
($3\times10^4$\,$L_\odot$). Taking into account that the star-forming region is
associated with at least one other hot molecular core (core A), it is unlikely
that its luminosity is only produced by core B.  This problem could be solved if the luminosity of the region has been 
underestimated because part of the stellar radiation escapes through the outflow cavities (Zhang et al.~\cite{zhang13}). Alternatively,  
 the existence of a binary system associated with core B could also be a solution (S\'anchez-Monge et
al.~\cite{sanchez-monge13}) because in this case, the total luminosity of the members of the system could be significantly 
lower than that of the whole star-forming region.  

  Our high-angular resolution 1.3\,cm and 7\,mm observations appear to support
  the second scenario because we have resolved the
centimeter emission associated with core B and confirmed 
the existence of a binary system  by
revealing two radio sources,  8a and 8b
(see Fig.~\ref{fig-disk}). The separation of the two sources
is $\sim$$0\farcs37$, which corresponds to $\sim$800~au at the distance of the
region.  Typical binary separations for T\, Tauri stars closely associated with
early-type stars is 90~au (Brandner \& Koehler~\cite{bradner98}). However, massive stars 
have companions with larger separations, in the range of
10$^3$ to several 10$^4$\,au (Sana \& Evans~\cite{sana11} and references
therein).

Source 8a is the strongest of the pair and could be the one associated with
the hot molecular core. Moreover, this source, as shown in
Fig.~\ref{fig-zoom}, is located closer to the center of the hot molecular core
seen in CH$_3$CN. If we assume that the free-free emission of source 8a is due to
an \UCHC\ region, then its spectral type would be B1 and its mass should be 
$\gtrsim$11\,$M_\odot$, as discussed in Sect.~\ref{nature}. The expected
luminosity for a B1 star is $\sim$6.3$\times10^3$\,$L_\odot$. In order to
account for the 18\,$M_\odot$ estimated by S\'anchez-Monge et
al.~(\cite{sanchez-monge13}) from the velocity field of the disk, source 8b
should have a mass $\lesssim8$\,$M_\odot$. The shape of the continuum spectrum 
of this source (Fig.~\ref{fig-index}), although uncertain, suggests that
the emission could be consistent with optically thin free-free emission (see
Sect.~\ref{spectral}). Therefore, we decided to estimate the spectral type from
the 1.3\,cm observations in the A configuration (because the emission of source 8b is
resolved from that of 8a) assuming that the centimeter emission arises from
an optically thin \UCHC\ region (see Table~\ref{table_uc}). The spectral type of source 8b would be B3,
which corresponds to a $\sim$6\,$M_\odot$ star with a luminosity of
$\sim$$10^3$\,$L_\odot$. This implies that the total mass of the binary system 
is $\gtrsim$17\,$M_\odot$ in agreement with the stellar mass estimated from the
velocity field, while its luminosity is $\sim$7$\times10^3$\,$L_\odot$, lower
 than $\sim$3$\times10^4$\,$L_\odot$, the luminosity of the whole
star-forming region.

In conclusion, our observations have revealed the presence of a binary system of
\UCHC\ regions at the center of a Keplerian rotating disk in \g35 (Fig.~\ref{fig-disk}). The existence of
a binary system solves the luminosity problem raised by S\'anchez-Monge et al.~(\cite{sanchez-monge13}), without invoking the
``flashlight effect''  as suggested by Zhang et al.~(\cite{zhang13}).
In this scenario, the
Keplerian disk discovered by S\'anchez-Monge et al.~(\cite{sanchez-monge13})
would indeed be a circumbinary disk, analogous to that observed in the young
low-mass binary system GG Tau (e.g., Skrutskie et al.~\cite{skrutskie93}).  
This close association of
the young binary system with a Keplerian disk is suggestive of a
scenario where the secondary 6~$M_\odot$ star source 8b has formed from disk
instabilities in an accretion disk around source 8a, which was initially
massive enough to become non-axisymmetric. The mass ratio (the secondary
to the primary) of this binary system is $\sim$0.54. Three-dimensional
radiation-hydrodynamic simulations by Krumholz et al.~(\cite{krumholz09}) demonstrate the formation of 
secondary massive stars arising from
disk instabilites with an initial mass ratio $>$0.5. This is a result
of the SWING mechanism in nearly Keplerian disks, described by Adams et al.
~(\cite{adams89})  and then simulated by Laughlin \&
Bodenheimer~(\cite{laughlin94}).

\subsection{Expansion of the radio jet}
\label{expansion}

\begin{figure*}
\centerline{
\hspace*{1.8cm}
\includegraphics[angle=-90,width=18cm]{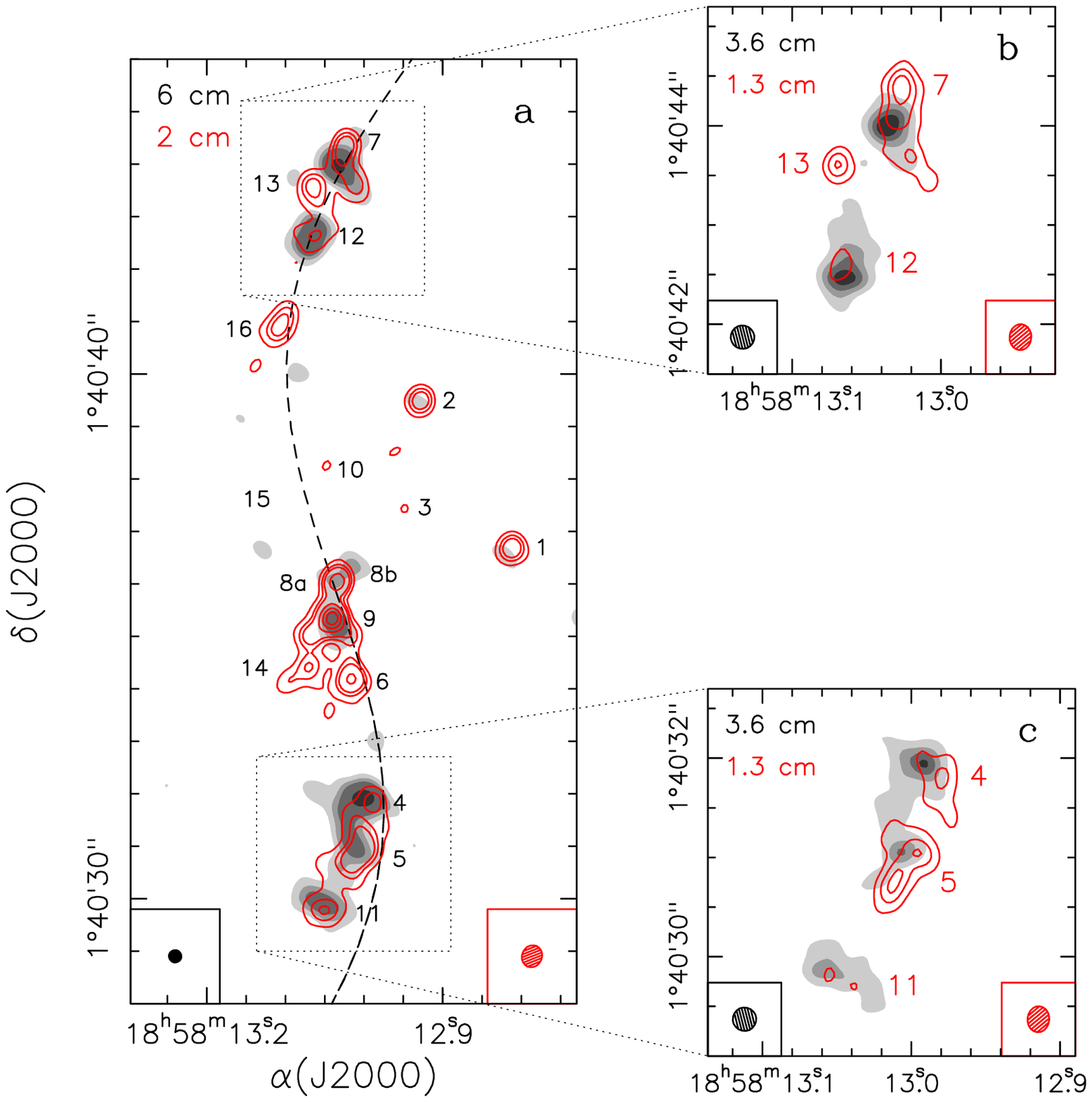}}
\caption{{\it a)} Overlay of the 2\,cm continuum emission ({\it red contours}) 
observed in B configuration on the 6\,cm continuum emission ({\it grayscale}) in
A configuration from Gibb et al.~(\cite{gibb03}).
Red contours are the same as in Fig.~\ref{fig-cm}a.  The grayscale levels are 
1, 2, 3, and 6 times 0.1\mjy, the 3\,$\sigma$ level of the map. The VLA 6 and 2\,cm synthesized beam are shown in the lower left and lower right 
corner, respectively. The numbers correspond to the centimeter continuum sources in 
Table~\ref{table_cont2}.  The black dashed line indicates the possible trajectory of the wiggling jet.
{\it b)} and {\it c)} Overlay of the 1.3\,cm continuum emission ({\it red contours}) 
observed in B configuration on the 3.6\,cm continuum emission ({\it grayscale}) in
A configuration from Gibb et al.~(\cite{gibb03}).
Red contours are the same as in Fig.~\ref{fig-cm}b.  The grayscale levels are 
1, 2, 3, and 4 times 0.08\mjy, the 3\,$\sigma$ level of the map. The VLA 3.6 and 1.3\,cm synthesized beam are shown in the lower left and lower right 
corner, respectively.}
\label{fig-compare}
\end{figure*}

Figure~\ref{fig-compare} shows the comparison of the radio jet emission at 
two different epochs, 1999 and 2013. The former is the epoch of the Gibb et
al.~(\cite{gibb03}) observations and the latter is that of our observations. 
This figure suggests that the radio jet is expanding. In fact, it shows that the radio knots at the northern and southern ends of the
jet appear to have moved outward with time because the emission of the knots
observed more recently, in particular at 1.3\,cm, is located ahead of the
emission of those observed by Gibb et al.~(\cite{gibb03}) at 6 and 3.6\,cm. Vice
versa, the position and compact morphology of the central radio sources have
not changed significantly with time, consistent with the fact that they are \UCHC\
regions.   We note that some of the southern knots show slightly transverse motions with respect to the 
direction of the jet (Fig.~\ref{fig-compare}c), similar to what has been observed towards the radio jet in 
IRAS~16547$-$4247 (Rodr\'{\i}guez et al.~\cite{rodriguez08}).

The difference between the old and the new positions of the knots is 
$\sim$$0\farcs40$ for source 7 in the north, and  $0\farcs20$--$0\farcs25$
for sources 4 and 5 in the south. These angular separations correspond to
$\sim$880\,au and 440--550\,au, respectively. For a  time lapse of 14 years, the
separation of the knots would indicate a velocity of the jet on the plane of the
sky of 150--300\,\kms, consistent with the proper motions measured for low-mass
radio jets (e.g., Anglada~\cite{anglada96}). Because these velocities are on the
plane of the sky, the real speeds may be higher. The proper motions of the
H$_2$O masers observed southward of source 8a also suggest that the jet is
expanding. In particular, the southernmost redshifted maser spot is expanding in
the same direction as that suggested by the free-free emission, namely S--SW.
However, the velocities of expansion inferred from the centimeter continuum
emission and the maser emission are different. The 3-D velocities of the
masers, $\lesssim50$~\kms, are much lower than the expansion velocities, which are
150--300\,\kms\ (at least), estimated from the radio jet emission. This is especially evident 
at the southern 
position where the maser emission coincides with the free-free knot 5. Here the 
expansion velocity of the jet (150--190~\kms) is 3--4 times
higher than the speed of the H$_2$O maser spot. 

Since both the free-free and maser emissions originate in shocks produced in the interaction of the
protostellar jet with circumstellar ambient material, the much lower velocities of the water masers 
can be explained in terms of the very high ($n_{\rm H_2}$ $\ge$ 10$^7$~\cmc) pre-shock density required to produce maser emission.
In fact, assuming that all the momentum of the jet is transferred to the shocked gas, it follows that, for a relatively light jet,
 the shock velocity depends inversely on the square root of the density
 of the ambient material (see, e.g., Masson \& Chernin~\cite{masson1993}).
Based on the  ratio of proper motion amplitudes of the radio knots and the water masers, we can infer that
the ionized gas emission should trace the interaction of the jet with gas parcels about one order of magnitude less dense
(i.e., $n_{\rm H_2}$ $\sim$ 10$^6$~\cmc) than those responsible for the water maser emission.

\subsection{Precession of the radio jet}
\label{precession}

As can see in Fig.~\ref{fig-compare}, the radio jet presents an S-shaped
morphology, which is strongly suggestive of precession, as has been found in
the case of other similar jets associated with low- and high-mass YSOs (e.g., 
HH--30: Anglada et al.~\cite{anglada07}; IRAS\,20126+4104: Shepherd et al.~\cite{shepherd00},
Cesaroni et al.~\cite{cesaroni05}). Precession is often due to the interaction between
the disk around the YSO powering the jet and a nearby stellar companion.

This provides us with the opportunity to tie the properties of the precessing
jet to the parameters of the binary system.  On the one hand, this
equation (from Shepherd et al.~\cite{shepherd00} and Terquem et al.~\cite{terquem99})
\begin{equation}
T_{\rm prec} = \frac{64\,\pi}{15} \frac{M_{\rm p}}{M_{\rm s}}
               \left(\frac{D}{R}\right)^3 \sqrt{\frac{R^3}{G\,M_{\rm p}}}
               \frac{\left(1-e^2\right)^{\frac{3}{2}}}{\cos\delta}
               \label{eqo}
\end{equation}
can be used to express the precession
period, $T_{\rm prec}$, as a function of the masses of the primary ($M_{\rm
p}$) and secondary ($M_{\rm s}$) members of the binary, the radius of the disk
($R$), the separation of the two stars ($D$), the eccentricity of the orbit
($e$), and the inclination of the binary orbit with respect to the plane of
the disk ($\delta$). On the other hand, $T_{\rm prec}$ can also be obtained
from the ratio between the wavelength ($\lambda$) of the sinusoidal-like
pattern of the jet and the jet expansion velocity ($v_{\rm jet}$):
\begin{equation}
T_{\rm prec}=\lambda/v_{\rm jet}.   \label{eqt}
\end{equation}

By combining Eqs.~(\ref{eqo}) and~(\ref{eqt}), the following
expression for $\lambda$ is obtained:
\begin{equation}
\lambda = \frac{64\,\pi}{15} v_{\rm jet} \frac{M_{\rm p}}{M_{\rm s}}
	       \frac{D^3}{\cos\delta\,\sqrt{G\,M_{\rm p}}}
	       \left(\frac{1-e^2}{R}\right)^{\frac{3}{2}}
	       \label{eqthree}
\end{equation}
This can be used to set a lower limit on $\lambda$.
In fact, the masses of the two stars are $M_{\rm
p}\simeq11~M_\odot$ and $M_{\rm s}\simeq6~M_\odot$ (see Sect.~\ref{binary}), while
the projected separation on the plane of the sky between the two
HC/UC~\HII regions 8a and 8b may be taken as a lower limit on $D$, i.e.
$D>800$~au. Moreover, 
$v_{\rm jet}\simeq200$~km~s$^{-1}$ (see Sect.~\ref{expansion}) and
$R\le D$ because the part of the disk affected by the companion lies inside
its orbit. Finally, $\cos\delta\le1$ and the eccentricity of the orbit is unlikely
to be very large and we thus assume $e<0.9$ (see Mathieu~\cite{mathieu94} and
Mason et al.~\cite{mason98}).
Taking all these values into account, we find $\lambda>0.45$~pc.

A value that large indicates that the radio emission is tracing only a small
fraction of the precession pattern, whose extent must exceed the region
shown in Fig.~\ref{fig-cm}. This finding is consistent with the scenario
proposed by S\'anchez-Monge et al.~(\cite{sanchez-monge14}).
By complementing the radio maps with near-IR images of the
2.12~$\mu$m H$_2$ line emission (see their Fig.~16), these authors propose
that the precessing jet/outflow from \g35 extends over a much larger region
than that traced by the radio jet itself. According to their interpretation,
the precession axis is not oriented N--S, but closer to a NE--SW direction,
and the value of $\lambda$ is higher than our previous
estimate. From Figs.~15 and~16 of S\'anchez-Monge et al.~(\cite{sanchez-monge14}), it 
is possible to roughly derive $\lambda/2\simeq30\arcsec$, i.e., $\lambda\simeq0.6$~pc, 
consistent with the lower limit derived above.

In conclusion, the properties of the binary system appear to support
precession of the jet with a relatively large precessing angle, about a
precession axis directed approximately NE--SW, which is in agreement with the
interpretation proposed by S\'anchez-Monge et al.~(\cite{sanchez-monge14}).

\section{Conclusions}

We carried out VLA observations in the continuum at 2 and 1.3\,cm, and 7\,mm  
of the high-mass star-forming region \g35 to study the origin of the free-free emission, to examine the binary system
hypothesis required to solve the luminosity problem in this source (S\'anchez-Monge et
al.~\cite{sanchez-monge13}), and to 
characterize the possible precession of the radio jet. The high angular resolution observations from 0.05$''$ to 0.45$''$ have allowed us to obtained
 a very detailed picture of the radio continuum emission in this region. The main results can be summarized as follows:

\begin{itemize}

\item We have detected 17 radio continuum sources in the region, some of them previously detected in the observations of Gibb et al.~(\cite{gibb03}). 
The spectral indices of the emission of the different sources are consistent with variable emission for most of them. For 4 out of 17 sources, the free-free emission is consistent with ionized emission from UC/HC HII regions.

\item The observations have revealed the presence of a binary system of  \UCHC\ regions 
at the geometrical center of the radio jet. This binary system, which is 
associated with the Keplerian  rotating disk discovered by S\'anchez-Monge et
al.~(\cite{sanchez-monge13}),  consists of two B-type stars of 11 and
6~$M_\odot$. The existence of a binary system solves the luminosity problem
raised by  S\'anchez-Monge et al.~(\cite{sanchez-monge13}) because the total
luminosity of the system would be $\sim$$7\times10^3$~$L_\odot$, much lower than
the luminosity of the whole star-forming region. The presence of such a
massive binary system  is in agreement with the theoretical predictions of
  Krumholz et al.~(\cite{krumholz09}), according to which the gravitational instabilities during the collapse
  would produce the fragmentation of the disk and the formation of such a system.

\item The 7 mm observations have detected 14 Class I CH$_3$OH maser spots in the region, most of them lying along the N--S direction of the radio jet 
and a few located in a N--E direction, coinciding with IR emission. This suggests an association between the methanol masers and the jet, consistent with the fact that Class I methanol masers are collisionally pumped.

\item The S-shaped morphology of the jet has been successfully explained as
being due to precession produced by the binary system located at its center.
By tying the properties of the precessing jet to those of the binary system,
we have set a lower limit of $\sim$0.45~pc on the typical wavelength of the
S-shaped pattern. This proves that the precession extends over a larger
region than that traced by the radio cm emission, consistent with a wide
precession angle around the NE--SW direction, as proposed by S\'anchez-Monge
et al.~(\cite{sanchez-monge14}).  In such a scenario, the three class~I CH$_3$OH maser spots located
to the east and west could be tracing a second flow.

 \item Comparison of the radio jet images obtained at 2 and 1.3\,cm in 2013 with those
obtained at 6 and 3.6 cm in 1999 (Gibb et al.~\cite{gibb03}) suggests that the
jet is expanding at a maximum speed on the plane of the sky of 300~\kms. The 
proper motions of the H$_2$O masers, detected in the region with the VERA interferometer, also indicate
expansion in a direction similar to that of the radio jet. 

\end{itemize}

\begin{acknowledgements}  

We thank the referee, Andr\'es Guzm\'an, for his useful comments. 
We acknowledge a partial support from the Italian Foreign Minister 
(MAECI) as a project of major importance in the Scientific and
Technological Collaboration between Italy and Japan.

 \end{acknowledgements}

\end{document}